\documentclass[superscriptaddress,dvipsnames,nofootinbib,amsmath,amssymb,aps,prd,floatfix,reprint]{revtex4-2}

\usepackage[titletoc,toc,title]{appendix}
\usepackage{mathtools}
\usepackage{amsmath, amsfonts,bm}
\usepackage[utf8]{inputenc}
\usepackage{xcolor}
\usepackage{graphicx}
\usepackage{float}
\usepackage{subfig}
\usepackage{comment}
\newcommand\myshade{85}
\colorlet{mylinkcolor}{RoyalPurple}
\colorlet{mycitecolor}{WildStrawberry}
\colorlet{myurlcolor}{BlueViolet}

\usepackage[normalem]{ulem}

\usepackage{hyperref}
\usepackage[capitalise]{cleveref}

\hypersetup{
  linkcolor  = mylinkcolor!\myshade!black,
  citecolor  = mycitecolor!\myshade!black,
  urlcolor   = myurlcolor!\myshade!black,
  colorlinks = true,
}

\newcommand\mathcomma{\,,}
\newcommand\mathperiod{\,.}

\DeclareMathAlphabet{\mathup}{OT1}{\familydefault}{m}{n}

\def\dd{\mathrm{d}}

\newcommand{\be}{\begin{equation}} 
\newcommand{\ee}{\end{equation}}

\usepackage{array}
\newcommand{\PreserveBackslash}[1]{\let\temp=\\#1\let\\=\temp}
\newcolumntype{C}[1]{>{\PreserveBackslash\centering}p{#1}}
\newcolumntype{R}[1]{>{\PreserveBackslash\raggedleft}p{#1}}
\newcolumntype{L}[1]{>{\PreserveBackslash\raggedright}p{#1}}
\renewcommand\arraystretch{1.5}
\usepackage{multirow}
\usepackage{bm}

\begin{document}

\title{Implications of distance duality violation for the $H_0$ tension and evolving dark energy}

\author{Elsa M. Teixeira}
\email{elsa.teixeira@umontpellier.fr}
\affiliation{Laboratoire Univers \& Particules de Montpellier, CNRS \& Université de Montpellier (UMR-5299), 34095 Montpellier, France}

\author{William Giar\`e}
\affiliation{School of Mathematics and Statistics, University of Sheffield, Hounsfield Road, Sheffield S3 7RH, United Kingdom}

\author{Natalie B. Hogg}
\affiliation{Laboratoire Univers \& Particules de Montpellier, CNRS \& Université de Montpellier (UMR-5299), 34095 Montpellier, France}

\author{Thomas Montandon}
\affiliation{Laboratoire Univers \& Particules de Montpellier, CNRS \& Université de Montpellier (UMR-5299), 34095 Montpellier, France}

\author{Adèle Poudou}
\affiliation{Laboratoire Univers \& Particules de Montpellier, CNRS \& Université de Montpellier (UMR-5299), 34095 Montpellier, France}

\author{Vivian Poulin}
\affiliation{Laboratoire Univers \& Particules de Montpellier, CNRS \& Université de Montpellier (UMR-5299), 34095 Montpellier, France}


\begin{abstract}

    We investigate whether a violation of the distance duality relation (DDR), $D_L(z) = (1+z)^2 D_A(z)$, connecting the angular diameter and luminosity distances, can explain the Hubble tension and alter the evidence for dynamical dark energy in recent cosmological observations. We constrain five phenomenological parameterisations of DDR violation using Baryon Acoustic Oscillation measurements from the DESI survey calibrated with the sound horizon derived from \textit{Planck} Cosmic Microwave Background data and the Pantheon+ Type Ia supernova (SNIa) catalogue calibrated with the supernova absolute magnitude from S$H_0$ES. We find that two toy models can resolve the tension: a constant offset in the DDR (equivalent to a shift in the calibration of the SNIa data), $D_L(z)/D_A(z)\simeq 0.925(1+z)^2$, which leaves the hint for evolving dark energy unaffected; or a change in the power-law redshift-dependence of the DDR, restricted to $z\lesssim 1$, $D_L(z)/D_A(z)\simeq(1+z)^{1.866}$, together with a {\it constant} phantom dark energy equation of state $w\sim -1.155$. The Bayesian evidence slightly favours the latter model. Our phenomenological approach motivates the investigation of physical models of DDR violation as a novel way to explain the Hubble tension.

\end{abstract}

\maketitle

\section{Introduction} \label{sec:intro}

The canonical $\Lambda$-cold-dark-matter ($\Lambda$CDM) model, in which dark energy is described by a cosmological constant $\Lambda$ and dark matter behaves as a dust-like component, has been remarkably successful in explaining a wide range of high-precision cosmological observations, including (but not limited to) the cosmic microwave background (CMB) radiation~\cite{Planck:2018vyg,ACT:2020frw,ACT:2023kun,SPT-3G:2021eoc}, baryon acoustic oscillations (BAO)~\cite{eBOSS:2020lta,eBOSS_cosmo}, and the apparent magnitude of type Ia supernovae (SNIa)~\cite{Brout:2021mpj}.

However, over the past decade, anomalies have emerged which raise questions about the validity of the $\Lambda$CDM model. There are indications that additional physical ingredients may be needed to restore concordance between datasets~\cite{Abdalla:2022yfr}. The most significant challenge to $\Lambda$CDM is the `Hubble tension', a $\sim 5\sigma$ discrepancy between the direct measurement of the Hubble parameter $H_0$ -- which measures the expansion of the Universe today -- using Cepheid-calibrated SNIa from the Supernovae $H_0$ for the Equation of State (S$H_0$ES) programme~\cite{Riess:2021jrx} and the $\Lambda$CDM prediction calibrated with CMB data measured by the {\it Planck} satellite~\cite{Planck:2018vyg}. Systematic errors in $H_0$ estimates are actively being searched for, \textit{e.g.}~\cite{Efstathiou:2020wxn,Mortsell:2021nzg,Mortsell:2021tcx,Riess:2021jrx,Sharon:2023ioz,Murakami:2023xuy,Riess:2023bfx,Bhardwaj:2023mau,Brout:2023wol,Dwomoh:2023bro,Uddin:2023iob,Riess:2024ohe,Freedman:2024eph,Riess:2024vfa}, and solutions to the tension which propose new physics have also been suggested~\cite{DiValentino:2021izs,Schoneberg:2021qvd,Abdalla:2022yfr,DiValentino:2025sru}.

This tension can be traced to a mismatch in the {\it calibration} of the cosmological distance indicators: the SNIa and the BAO observed in various tracers of the density field. When both the S$H_0$ES absolute magnitude calibration and the {\it Planck} $\Lambda$CDM sound horizon calibration are applied, a substantial discrepancy arises between the SNIa and BAO data~\cite{Pogosian:2021mcs,Tutusaus:2023cms,Raveri:2023zmr,Bousis:2024rnb,Poulin:2024ken}. Under the $\Lambda$CDM model, this tension results in values for $H_0$ derived from the two datasets that are incompatible at $5\sigma$. This incompatibility is also reflected in other parameters, in particular the physical matter density $\Omega_{\rm m}h^2$ (where $\Omega_{\rm m}$ is the fractional density of baryons and cold dark matter in the Universe and $h\equiv H_0/(100\, \text{km}\, \text{s}^{-1}\, \text{Mpc}^{-1})$) ~\cite{Jedamzik:2020zmd,Blanchard:2022xkk,Poulin:2024ken,Pedrotti:2024kpn}, and is known as the `cosmic calibration tension'. Moreover, even in the absence of the S$H_0$ES calibration, hints of deviations from $\Lambda$CDM have appeared when combining BAO and SNIa data, with a preference for dynamical dark energy at the $3-4\sigma$ level depending on the exact datasets used ~\cite{Rubin:2023ovl, DES:2024jxu, DESI:2024mwx,DESI:2025zgx}. 

From a more fundamental perspective, these tensions can be seen as a tension between two sets of cosmic distance measurements. Cosmological observations 
infer either the \textit{angular diameter distance}, $D_A(z)$, from the apparent angular size of an object of known physical size, or the \textit{luminosity distance}, $D_L(z)$, from the flux of a source of known intrinsic brightness. The cosmic calibration tension arises when comparing luminosity and angular diameter distances. However, this comparison is made with the (often implicit) assumption that the distance-duality relation (DDR)~\citep{Etherington1933,Ellis2007}, $D_{L}(z) = (1+z)^2 D_{A}(z)$, holds in our Universe, allowing for direct conversion between $D_A(z)$ and $D_L(z)$. 

The DDR holds in any cosmological model that satisfies the following conditions: i) spacetime is described by a pseudo-Riemannian manifold, ii) photons propagate along (unique) null geodesics, and iii) their number is conserved over time. DDR violations can range from the prosaic, such as photon scattering off dust or free electrons, to the exotic~\cite{Ellis:2013cu,Bassett:2003vu}. They can emerge in modified theories of gravity where spacetime is not described by a pseudo-Riemannian manifold~\cite{Santana:2017zvy,Azevedo:2021npm}, such as in torsion-based theories, where the connection is not purely Levi-Civita but includes torsional terms~\cite{Cai:2015emx}, and in energy--momentum squared gravity, where the field equations include higher-order terms of the energy-momentum components~\cite{Cipriano:2024jng}. DDR violation can further occur in scenarios beyond the Standard Model where photons acquire an effective mass through interactions with axion-like particles or where their number density decreases through decay into these particles~\cite{Avgoustidis:2010ju,Csaki:2001yk,Bassett:2003zw}. 
Yet further possibilities include photons coupling to gravitons in an external magnetic field~\cite{Chen:1994ch,Cillis:1996qy,Raffelt:1987im}; dust extinction or by interstellar gas attenuation \cite{Menard:2009yb}; Kaluza--Klein models associated with extra dimensions~\cite{Deffayet:2000pr}; chameleon fields with couplings to matter of order unity~\cite{Khoury:2003aq,Khoury:2003rn}; photon--chameleon mixing~\cite{Burrage:2007ew}; and phenomenological scenarios where fundamental constants, such as the speed of light~\cite{Barrow:1999is,Lee:2021xwh} or the fine structure constant~\cite{Goncalves:2019xtc} vary over time. 

These physical processes can be constrained by current data\footnote{For works spanning the past two decades discussing the possibility of testing the DDR against a multitude of astrophysical and cosmological datasets in a model-dependent and -independent manner, we refer to Refs.~\cite{Uzan:2004my,DeBernardis:2006ii,Holanda:2010ay,Holanda:2010vb,Li:2011exa,Nair:2011dp,Liang:2011gm,Meng:2011nt,Holanda:2011hh,Khedekar:2011gf,Goncalves:2011ha,Lima:2011ye,Holanda:2012at,Cardone:2012vd,Yang:2013coa,Zhang:2014eux,Santos-da-Costa:2015kmv,Wu:2015ixa,Wu:2015ixa,Liao:2015uzb,Rana:2015feb,Ma:2016bjt,Holanda:2016msr,More:2016fca,Rana:2017sfr,Li:2017zrx,Lin:2018qal,Qi:2019spg,Holanda:2019vmh,Zhou:2020moc,Qin:2021jqy,Bora:2021cjl,Mukherjee:2021kcu,Liu:2021fka,Renzi:2021xii,Tonghua:2023hdz,Qi:2024acx,Yang:2024icv,Tang:2024zkc,Jesus:2024nrl,Qi:2024acx,Alfano:2025gie,Yang:2025qdg,Keil:2025ysb,Xu:2022zlm}. For projected limits expected from future surveys, see, \textit{e.g.}, Refs.~\cite{Cardone:2012vd,Yang:2017bkv,Fu:2019oll,Hogg:2020ktc,Renzi:2020bvl,Euclid:2020ojp}.}, but the constraining power of the data on a given model depends on the specifics of the model and the datasets in question. Nevertheless, if \textit{any} violation of the DDR occurs in nature, it could affect the calibration of distance indicators, with implications ranging from affecting the $\sim 5\sigma$ $H_0$-tension in $\Lambda$CDM to potentially influencing the preference for a dynamical dark energy component.

In this paper, we explore three questions related to the violation of the DDR:

\begin{enumerate}
    \item Can simple phenomenological parameterisations of DDR violation in a $\Lambda$CDM background cosmology reduce or eliminate the tension between calibrated SNIa and BAO?
    \item Is there evidence for a violation of the DDR that changes over cosmic history?
    \item Can a DDR violation alter the preference for dynamical dark energy observed in the combination of current BAO and SNIa data?
\end{enumerate}

A violation of the DDR that is constant in time is mathematically identical to a mismatch in the calibration of luminosity and angular diameter distance measurements. Evidence for a deviation from the DDR that changes in time would suggest that models proposed to resolve the calibration tension and that do so {\it solely} by altering the sound horizon or the SNIa magnitude (including systematic errors) are disfavoured. Therefore, examining the preference for a redshift-dependent DDR violation can be viewed as a crucial null test for new physics. Concretely, we find that allowing for a phenomenological violation of the DDR can indeed resolve the calibration tension between SNIa and BAO. In fact, we find two possibilities currently favoured by the data. 
A constant violation of the DDR (equivalent to a calibration shift), $D_L(z)/D_A(z)\simeq 0.925(1+z)^2$; or a change in the power-law redshift-dependence of the DDR, restricted to $z\lesssim 1$, $D_L(z)/D_A(z)\simeq(1+z)^{1.866}$, together with a phantom dark energy equation of state $w\sim -1.155$ (similar to Ref.~\cite{Tutusaus:2023cms}). The data slightly favour the latter scenario over the former, suggesting a novel way to address the Hubble tension.
Our approach is phenomenological; we leave the development of a physical model of DDR breaking that would produce such an effect to future work.

This paper is structured as follows: in \cref{sec:ddrtheory}, we review the DDR and the calibration of distance measurements. In \cref{sec:method}, we discuss our methodology, focusing on the different parameterisations used to test potential violations of the DDR, the datasets involved, and the statistical techniques employed to derive our results. In \cref{sec:results_lcdm}, we present the main findings of our analysis regarding the Hubble tension, answering questions 1 and 2 above. In \cref{sec:DDE}, we investigate the preference for evolving dark energy in the face of DDR violation, answering question 3 above. We present further discussions and conclusions in \cref{sec:discussion} and \cref{sec:conclusions}. Throughout this paper, we work in units such that $8\pi G = c = 1$. 

\section{The DDR, distance measurements and cosmology} \label{sec:ddrtheory} 

The DDR states that~\citep{Etherington1933,Ellis2007}
\begin{equation}
\label{eq:DDR}
    D_{L}(z) = (1+z)^2 D_{A}(z) \mathcomma
\end{equation}  
where $D_{L}(z)$ is the \textit{luminosity distance} and $D_{A}(z)$ is the \textit{angular diameter distance}. Our analysis assumes that the cosmological principle holds -- \textit{i.e.}, that the Universe is homogeneous and isotropic on sufficiently large scales. Consequently, we do not consider any directional dependence of the DDR but instead focus on its possible redshift dependence.\footnote{For further discussions on consistency tests and violations of the cosmological principle, see \textit{e.g.} \cite{Clarkson:2010uz,Clarkson:2011br,Maartens:2011yx,Ntelis:2017nrj,Aluri:2022hzs}.} Under these assumptions, measurements of luminosity and angular diameter distances at different redshifts can be used to test the DDR. These tests can be performed either by considering specific physical models that predict DDR violations (hence constraining the models using data) or in an empirical model-agnostic way by rewriting the DDR as  
\begin{equation}
    \eta(z) = \frac{D_{L}(z)}{(1+z)^2 D_{A}(z)} \mathcomma \label{eq:eta_ddr}
\end{equation}  
where deviations from $\eta(z) = 1$ indicate a violation of the DDR.  

We follow the latter approach, using SNIa as our probe of luminosity distances and BAO data as our probe of angular diameter distances to test for violations of $\eta(z) = 1$ using different phenomenological parameterisations of the function $\eta(z)$. The specific datasets and parameterisations used are detailed in~\cref{sec:method}. Next, we describe how luminosity and angular diameter distances are obtained from SNIa and BAO.

\subsection{Distances in cosmology} \label{sec:dl}

The luminosity distance, as measured from SNIa, is given by 
\begin{align}
     5 \log_{10} {D_{L} (z)/ \text{Mpc} } &= \mu(z)  + 5  \,\nonumber\\ 
      &= m(z) - M_B  + 5 \mathcomma
    \label{eq:dl_sn}
\end{align}
where we have defined the distance modulus $\mu = m(z) -M_B$, $m(z)$ is the apparent magnitude of the SNIa obtained from their light curves, and $M_B$ is the absolute magnitude of the SNIa, which is not known from first principles and must be calibrated for, using objects whose intrinsic luminosity is known. Thanks to their well-measured period-luminosity relation, Cepheid variable stars have historically played this role. Calibrating the S$H_0$ES SNIa builds a distance ladder involving four geometric anchors to calibrate the period-luminosity relation of Cepheids \cite{Riess:2021jrx}. The calibration of the Cepheids is used to calibrate the distances to 42 SNIa observed in 37 galaxies that also host Cepheids. These 42 SNIa are then used to calibrate the distances to SNIa observed in galaxies in the Hubble flow that can be found within compilations such as Pantheon+ (or Pantheon-Plus) \cite{Brout:2021mpj}.

On the other hand, the BAO observed within various tracers of the matter density field (\textit{e.g.} galaxies, quasars or the Lyman-$\alpha$ forest) can be used to measure the angular diameter distance via 
\begin{equation}
    D_A (z) = \frac{r_s(z_d)}{\theta_s(z)} \mathcomma
    \label{eq:theta_bao}
\end{equation}
where $\theta_s(z)$ is the angular size of the BAO feature and $r_s(z_d)$ is the physical size of the sound horizon at the redshift of baryon drag $z_d$,
\begin{equation}
    r_s(z_{d}) = \int_{z_d}^{\infty} \frac{c_s(z)}{H(z)} \, \dd z \mathcomma
\end{equation}
and $c_s\simeq 1/\sqrt{3}$ is the sound speed in the tightly coupled photon--baryon plasma. The Hubble rate $H(z)$ describes the expansion history of the Universe.

Since galaxy surveys have access to 3D information, they can also be used to measure the longitudinal (along the line of sight) BAO,
\begin{equation}
\label{eq:DH}
D_H(z)=\frac{r_s}{\Delta z}= \frac{1}{H(z)} \mathcomma
\end{equation} 
which allows a direct measurement of $H(z)$ to be derived at the redshift of the survey. Furthermore, for measurements with low signal-to-noise ratio, it is common to build the angle-average distance $D_V(z)$,
\begin{equation}
    D_V(z)\equiv \left[zD_M(z)^2D_H(z)\right]^{1/3}\mathperiod
\end{equation}
These data are crucial to breaking the degeneracy between the signatures of a different expansion history $H(z)$ and the effect of a DDR violation.

\subsection{From distances to cosmology}

Beyond tests of the DDR, distance measurements are routinely used to constrain cosmological models. Assuming a homogeneous and isotropic Universe described by the flat Friedmann--Lema\^itre--Robertson--Walker metric allows us to write the angular diameter distance as 
\begin{equation}
\label{eq:DA_cosmo_bis}
    D_A(z)\equiv\frac{H_0}{(1+z)} \,\int_0^{z'}\frac{\mathrm{d}z'}{H^2(z')} \mathperiod
\end{equation}

In this work, we consider two different baseline cosmologies for which the expansion history is given by 
\begin{multline}
    \frac{H^2}{H_0^2} = \Omega_{\rm r}(1+z)^4+\Omega_{\rm m}(1+z)^3 \vphantom{\Omega_{\rm DE}\,e^{3 \int_0^z \frac{1+w(z')}{1+z'} \dd z'}}\\
     +\; \Omega_{\rm DE}\,\exp \left[3 \int_0^z \frac{1+w(z')}{1+z'} \dd z' \right] \mathcomma 
\end{multline}
with $\Omega_{\rm r}$, $\Omega_{\rm m}$ and $\Omega_{\rm DE}$ respectively being the radiation, matter and dark energy density today, and where $w(z)$ is the dark energy equation of state.

We first consider the flat $\Lambda$CDM model, corresponding to the special case $w(z)\equiv-1$, in which $\Omega_{\rm DE}$ describes a cosmological constant $\Lambda$. We then investigate the so-called $w_0w_a$CDM scenario in which the late-time accelerated expansion of the Universe is driven by a perfect fluid with equation of state given by the Chevalier--Polarski--Linder (CPL) parameterisation (also known as $w_0w_a$ parameterisation)~\cite{Chevallier_Polarski_2001, Linder:2002et},
\begin{equation}
\label{eq:CPL}
    w(a) = w_0 + (1-a) w_a \mathcomma
\end{equation}
where $a\equiv 1/(1+z)$ is the cosmic scale factor. The cosmological constant is recovered when $w_0 = -1$ and $w_a = 0$. 

\subsection{A sign of DDR violation?}
 
\begin{figure*}
      \includegraphics[width=\textwidth]{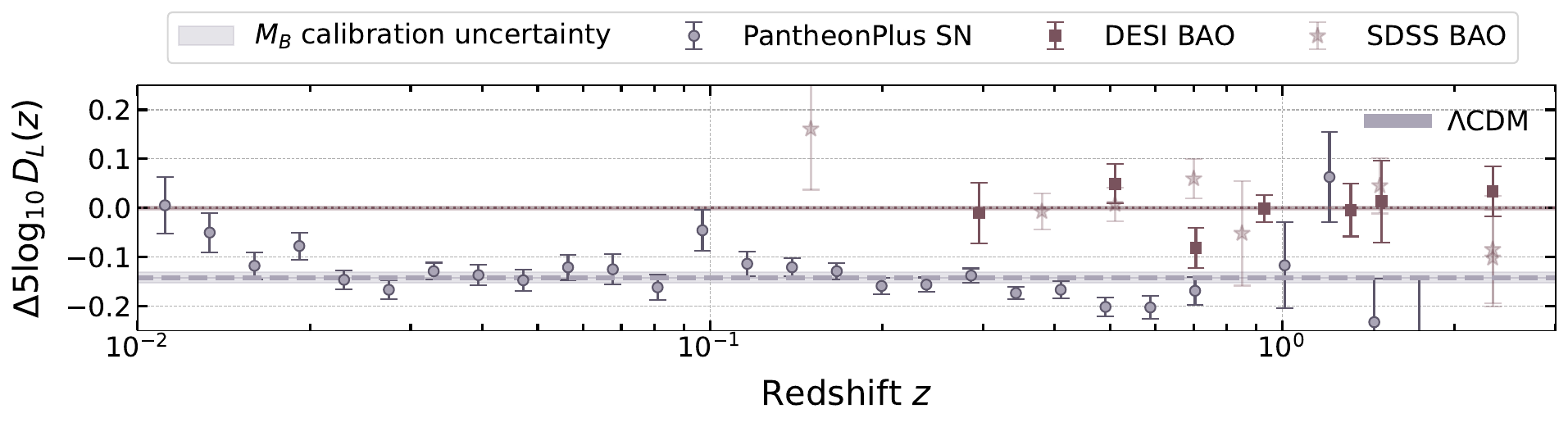}
  \caption{\label{fig:sn_bao_lcdm_only} Mismatch between SNIa and BAO data under a $\Lambda$CDM background through the combination of \textit{Planck} 2018 + DESI + PantheonPlus + S$H_0$ES prior data as described in \cref{sec:data}. The square data points represent a compilation of angular diameter distance measurements from DESI BAO data (the star data points show an alternative BAO sample from the completed SDSS-IV eBOSS survey \cite{BOSS:2016wmc,eBOSS:2020lta,eBOSS:2020hur,eBOSS:2020qek,eBOSS:2020fvk,eBOSS:2020gbb,eBOSS:2020uxp,eBOSS:2020tmo,Ross:2014qpa,Howlett:2014opa,eBOSS:2020yzd}). The round data points show the difference between the distance moduli $\mu$, defined in \cref{eq:dl_sn}, of the Pantheon$+$ SNIa catalogue, calibrated with S$H_0$ES. We note that the observed offset reflects the incompatibility between the datasets expressed as different predictions for $M_B$.}
\end{figure*}

Whilst tests of the DDR can be made without specifying a value for the absolute magnitude of the SNIa, the Hubble tension brought about by the S$H_0$ES calibration leads to a sign of potential DDR violation. We illustrate this in \cref{fig:sn_bao_lcdm_only}, where we show the difference between the distance modulus $\mu(z)$, see \eqref{eq:dl_sn}, of SNIa in the Pantheon+ catalogue calibrated with S$H_0$ES (grey points) and the transverse BAO sample from the DESI collaboration (brown squares) \cite{DESI:2024uvr,DESI:2024lzq} and BOSS collaboration (light brown stars) \cite{eBOSS:2020lta,Alam:2016hwk} in the best-fit $\Lambda$CDM cosmology derived from a combination of \textit{Planck} CMB, (uncalibrated) Pantheon+ and DESI data.\footnote{We followed the approach described in App.~A of Ref.~\cite{Poulin:2024ken} to bin the Pantheon+ catalogue of 1550 SNIa.} These results are obtained by applying the DDR to translate the angular diameter distance into the distance modulus. The mismatch between SNIa and BAO data when the S$H_0$ES and {\it Planck} $\Lambda$CDM calibrations are imposed simultaneously is clearly apparent.
As we study in the rest of this paper, a violation of the DDR would translate, or `move' the grey points upwards on the y-axis, potentially bringing them into better agreement with the brown points.
However, the DDR cannot alter the BAO points since BAO also measures longitudinal information, $D_H(z)$, which is unaffected by the DDR. 

\section{Methodology} \label{sec:method}
\subsection{Parameterising the DDR violation} \label{sec:param}
We choose to explore a set of simple parameterisations that can capture the phenomenology of DDR violation due to real physical models without assuming a specific theory. Whilst phenomenological models may be subject to additional constraints not considered here, this approach provides a \textit{proof-of-principle} test of whether a given toy-model parameterisation can resolve the Hubble tension or offer an alternative explanation for the preference for dynamical dark energy over a cosmological constant. Our parameterisations are listed below. We refer to them as M1, M2, M3, M1($z_*$) and M3($z_*$) throughout the text.

\begin{itemize}
\item \textbf{M1 -- constant DDR violation:} 
\begin{equation}
\eta(z) = 1 + \alpha_0 
\label{Eq:M1}
\end{equation}
In this case, only one additional parameter, $\alpha_0$, appears in \cref{eq:eta_ddr}. This parameter can be directly mapped to a rescaling of the sound horizon,
\begin{equation}
\label{eq:rs_rescale}
     1+\alpha_0=\frac{r_s^{\rm SH0ES}}{r_s^{\Lambda \rm CDM}}\mathcomma
\end{equation}
where $r_s^{\Lambda \rm CDM}$ represents the value of the sound horizon obtained from the fit of $\Lambda$CDM to {\it Planck} data, while $r_s^{\rm SH0ES}$ is the sound horizon required to fit the combined BAO+SNIa+S$H_0$ES dataset~\cite{Bernal:2016gxb,Aylor:2018drw}. One can also map $\alpha_0$ to a change in the SNIa magnitude:
\begin{equation}
\label{eq:Mb_rescale}
5\log_{10}(1+\alpha_0)=M_B^{\Lambda \rm CDM}-M_B^{\rm SH0ES}\mathcomma
\end{equation}
where $M_B^{\Lambda \rm CDM}$ is the value obtained when fitting \textit{Planck} + SNIa data under $\Lambda$CDM, and $M_B^{\rm SH0ES}$ is the S$H_0$ES value. 

This model should be understood as a reference case that is mathematically equivalent to altering the calibration of the distance measurement data rather than capturing realistic physical violations of the DDR. We aim to test whether any more complex model can perform better (in a Bayesian sense) than this simple calibration-like extension.

\item \textbf{M2 -- linear DDR violation:} 
\begin{equation}
\eta(z) = 1 + \alpha_0 + \alpha_1 z
\label{Eq:M2}
\end{equation}   
The linear model defined by \cref{Eq:M2} (hereafter referred to as M2) introduces two additional parameters: $\alpha_0$ and $\alpha_1$. This model can be viewed as a first-order Taylor expansion of $\eta(z)$ at low redshift. In this interpretation, the constant term of the expansion of $\eta(z)$ at $z=0$ is split into the form $1 + \alpha_0$ to explicitly isolate any constant-violating term $\alpha_0$, while $\alpha_1 \equiv \frac{d\eta}{dz}$ at $z=0$ quantifies the strength of the redshift evolution of the function. The DDR is recovered when both $\alpha_0$ and $\alpha_1$ vanish. 

\item \textbf{M3 -- power-law DDR violation:}
\begin{equation}
\eta(z) = (1+z)^{\alpha_0}
\label{Eq:M3}
\end{equation}  
The power-law model defined by \cref{Eq:M3} (hereafter referred to as M3) has only one free parameter, $\alpha_0$, and the DDR is recovered when $\alpha_0 = 0$. 
Regarding its physical interpretation, a power-law provides a natural framework for describing DDR violations that may arise from a violation of photon number conservation. For the case of CMB photons, this effect can be related to changes in the temperature-redshift relation $T(z)=T_0 (1+z)$.
As shown in Ref.~\cite{Euclid:2020ojp}, assuming for simplicity that such deviations are achromatic (\textit{i.e.}, independent of photon wavelength) and approximately adiabatic (so that the CMB radiation spectrum remains close to a blackbody), the resulting modification to the evolution of the CMB temperature with redshift can be parameterised as $T(z) = T_0(1+z)^{1-\beta}$. This induces a power-law-like break in the DDR that is well captured by M3, with $\alpha_0 = -\frac{3}{2} \beta$~\cite{Avgoustidis:2011aa,Luzzi:2009ae,Lima:2000ay}.
\end{itemize}

Note that in M1 and M3, the DDR-violating parameter $\alpha_0$ affects $\eta(z)$ either as an additive constant (M1) or in the exponent of the power law (M3). However, in physically motivated scenarios, deviations from the DDR may not be strictly constant but could instead vary with redshift. To account for this possibility, we explore a redshift-dependent extension of both M1 and M3. Specifically, following the approach of Ref.~\cite{Euclid:2020ojp}, we divide $\eta(z)$ into two redshift bins, allowing for a more flexible parameterisation of possible DDR violations. 
This leads to two additional models:  

\begin{itemize}
\item \textbf{M1}\bm{$(z_*)$} -- \textbf{constant with transition at} \bm{$z_{*}$}\textbf{:} 
\begin{equation}
\eta(z) = \begin{cases} 1+\alpha_0\quad z<z_{*}\mathcomma \\ 1+\alpha_1\quad z \geq z_{*}\mathperiod \end{cases}
\label{Eq:M1(z)}
\end{equation}
The model defined by \cref{Eq:M1(z)} is the redshift-dependent generalisation of M1, where the constant DDR violation can vary at different redshifts. In particular, $z_{*}$ defines a transition redshift, after which the magnitude of the constant DDR violation can change. For this reason, we label it M1($z_*$). Note that we have two free parameters, $\alpha_0$ and $\alpha_1$, which we will sample in our statistical analyses while keeping $z_*$ fixed within each run. This approach is akin to the study performed in \cite{Perivolaropoulos:2023iqj}, where authors let the SNIa take on distinct values of the absolute magnitude $M_B$ at different redshift bins.

\item \textbf{M3}\bm{$(z_*)$} --  \textbf{power-law with transition at} \bm{$z_{*}$}\textbf{:} 
\begin{equation}
\eta(z) = \begin{cases} (1+z)^{\alpha_0}\quad z<z_{*}\mathcomma \\ (1+z)^{\alpha_1}\quad z \geq z_{*}\mathperiod \end{cases}
\label{Eq:M3(z)}
\end{equation}
Analogous to M1($z_*$), the model defined by \cref{Eq:M3(z)} is the redshift-dependent generalisation of M3, where the exponent of the power-law DDR violation can differ before and after the transition defined by $z_{*}$. We label this model M3($z_*$). Again, we have two free parameters, $\alpha_0$ and $\alpha_1$, which we will sample in our statistical analyses while keeping $z_*$ fixed.

\end{itemize}

Throughout this study, we fix $z_* = 0.9$, which corresponds to taking approximately half of the BAO redshift range in each of the two bins. In \cref{appendix:redshift}, we report on the results considering seven different values for the separation redshift: $z_{*} = \{0.4, 0.6, 0.8, 0.9, 1.0, 1.4, 2.0\}$. These values are chosen so as to incrementally add/subtract one BAO data point from each redshift bin and assess the impact of taking different sub-samples of the full dataset.

\subsection{Analysis setup}

We implement our parameterisations of the function $\eta$ listed in \cref{sec:param} as a modification to the computation of the luminosity distance in the Einstein-Boltzmann solver \texttt{CLASS}\footnote{\href{https://github.com/lesgourg/class_public}{https://github.com/lesgourg/class\_public}}~\cite{lesgourgues2011cosmic,Blas_2011,lesgourgues2011cosmic2}, \texttt{CLASS\_DDR}\footnote{\href{https://github.com/elsateixeira/class_ddr}{https://github.com/elsateixeira/class\_ddr}} including the additional DDR parameters as input parameters. This is complemented by a new likelihood module in which the SNIa data are compared to theoretical predictions of the distance moduli in terms of the modified luminosity distance derived from the combined \cref{eq:dl_sn,eq:eta_ddr}. In contrast, the BAO data are compared to combinations of the Hubble parameter $H(z)$ and the standard angular diameter distance defined in \cref{eq:theta_bao}. 

We perform Bayesian inference using the publicly available package \texttt{MontePython}\footnote{\href{https://github.com/brinckmann/montepython_public}{https://github.com/brinckmann/montepython\_public}}~\cite{Brinckmann:2018cvx,Audren_2013}. We use the nested sampling algorithm provided by the MultiNest\footnote{\href{https://github.com/farhanferoz/MultiNest}{https://github.com/farhanferoz/MultiNest}} \cite{Feroz:2007kg,Feroz:2008xx,Feroz:2013hea} package interfaced by the PyMultiNest\footnote{\href{https://github.com/JohannesBuchner/PyMultiNest}{https://github.com/JohannesBuchner/PyMultiNest}} \cite{Buchner:2014nha} package. Our decision to use nested sampling rather than Metropolis--Hastings Markov chain Monte-Carlo (MCMC) is motivated by the large number of degeneracies between the background cosmological parameters ($H_0$ and $\Omega_m$), the calibrators ($M_B$ and $r_s$) and the DDR parameters ($\alpha_0$ and $\alpha_1$). Vanilla Metropolis--Hastings is not well-adapted to drawing samples from highly degenerate posteriors and can further become trapped in local minima. An additional advantage of nested sampling is that it allows direct computation of the Bayesian evidence, which we will use in our model comparison. We analyse the output of our nested sampling runs and produce contour plots using the \texttt{GetDist}\footnote{\href{https://github.com/cmbant/getdist}{https://github.com/cmbant/getdist}} Python package~\cite{Lewis:2019xzd}.

\begin{table}[htbp!]
\begin{center}
\begin{tabular}{cc}
Parameter                    & Prior \\
\hline
\hline
$\Omega_{\rm b} h^2$                & $[0.020,0.025]$ \\
$\Omega_{\rm cdm} h^2$                & $[0.1,0.15]$ \\
$h$                    & $[0.55,0.80]$ \\
$\tau_{\rm reio}$                          & $[0.02, 0.08]$ \\
$n_s$                      & $[0.93,1.00]$ \\
$\log 10^{10}\, A_{s}$   & $[2.9, 3.2]$ \\
\hline 
$w_0$ & $[-3,1]$ \\
$w_a$ & $[-3,2]$ \\
\hline 
$\alpha_0$ & $[-0.5,0.5]$ \\
$\alpha_1$ & $[-0.5,0.5]$ \\
\hline 
\end{tabular}
\end{center}
\caption[Priors on the model parameters]{Flat priors on the cosmological and model parameters sampled in this work using the combination of SNIa, BAO and CMB data as detailed in \cref{sec:data}.}
\label{tab:priors_planck}
\end{table}

We sample the standard $\Lambda$CDM parameters \{$\Omega_{\rm b} h^2$,$\Omega_{\rm cdm} h^2,h, \tau_{\rm reio}, n_s, A_{s} \}$, where $h = H_0 /100$ is the reduced Hubble parameter, $\Omega_{\rm b}$ and $\Omega_{\rm cdm}$ are the baryon and CDM energy densities today, respectively and $\tau_{\rm reio}$ is the integrated reionisation optical depth. The primordial power spectrum is assumed to follow a power law with amplitude $A_s$ at $0.05\,h/$Mpc and tilt $1-n_s$ where $n_s=1$ is the scale-invariant limit. 
Also included in our analysis are $M_B$, a nuisance parameter in the SNIa likelihood, and the nuisance parameter $A_{\text{Planck}}$ for the overall amplitude of the \texttt{Plik\_lite} \textit{Planck} likelihood for high multipoles. To model the DDR violation following the $\eta$-parameterisation, we additionally sample $\alpha_0$ or both $\{\alpha_0,\alpha_1\}$ according to the model under consideration. We firstly perform our analyses assuming that dark energy is a cosmological constant, $w(z)=-1$ and then repeat them assuming that dark energy can be dynamical, using the CPL parameterisation given in \cref{eq:CPL}, in which case we also sample the parameters $\{w_0,w_a\}$. Neutrinos (and other unspecified parameters) are treated in the same way as the \textit{Planck} collaboration \cite{Aghanim:2018eyx}. A summary of the parameters and the flat priors employed is given in \cref{tab:priors_planck}.

\subsection{Data} \label{sec:data}

To derive constraints on possible deviations from the standard DDR, we use a compilation of data from CMB, SNIa and BAO surveys, as detailed below:

\begin{itemize}
\item \textbf{\textit{Planck} 2018}: To calibrate the BAO, we use the high-$\ell$ TTTEEE CMB power spectrum data\footnote{Let us stress that there are realistic models of DDR violations that leave an imprint in the CMB that is not captured by our parameterisation, \textit{e.g.} by altering the temperature of CMB photons~\cite{Euclid:2020ojp}. We focus here on the phenomenological aspects in the context of the Hubble tension and thus do not consider those effects here.} from \textit{Planck} 2018, using the \texttt{Plik} likelihood~\cite{Planck:2018vyg}. We use the foreground-marginalised likelihood `\texttt{Plik\_lite}' to accelerate parameter inference~\cite{Planck:2015bpv}. We also include the PR3 low-$\ell$ TT and EE likelihoods~\cite{Planck:2018vyg,Planck:2018nkj,Planck:2019nip}. 

\item \textbf{Pantheon+ (SN)}: distance modulus measurements from 1701 light curves of 1550 spectroscopically confirmed SNIa in the redshift range $0.001 < z < 2.26$~\cite{Scolnic:2021amr,Brout:2022vxf}. The authors in \cite{Brout:2022vxf} present the post-processed observed magnitudes, including residuals for corrections for various systematic effects \cite{Brout:2021mpj} and marginalisation over the relevant nuisance parameters (see, \textit{e.g.}, \cite{Nordin:2008aa,Efstathiou:2024xcq,Dhawan:2024gqy} for an examination of the impact of systematics in cosmological parameter inference). 
    
\item \textbf{DESI BAO}: BAO measurements from the first year\footnote{During completion of this work, the DESI collaboration has published a new set of BAO measurements \cite{DESI:2025zgx}. These new data are statistically compatible with the older ones used in this work. We do not expect our conclusions to be affected by the newer data, except when assessing the preference for evolving dark energy, which has strengthened with the new data release.} of observations of the Dark Energy Spectroscopic Instrument (DESI). The data is derived from observations of galaxies, quasars~\cite{DESI:2024uvr}, and Lyman-$\alpha$ tracers~\cite{DESI:2024lzq}, as listed in Table I of Ref.~\cite{DESI:2024mwx}. The data covers a redshift range $z\sim0.1-4.1$, expressed as measurements of the transverse comoving distance $(D_M/r_d)$, the Hubble horizon $(D_H/r_d)$, and the angle-averaged distance $(D_V/r_d)$.\footnote{The comoving distance is related to the angular diameter distance via $D_M = D_A (1+z)$.} The correlations between measurements of $D_M/r_d$ and $D_H/r_d$ are taken into account. 

\item \textbf{S$H_0$ES}: We include the S$H_0$ES data through a Gaussian prior on the SNIa intrinsic magnitude $M_B = -19.253 \pm 0.027$ \cite{Riess:2021jrx}. This approach was found to yield identical results to the use of distance anchors through the full PantheonPlus+S$H_0$ES likelihood, see, \textit{e.g.}, Ref.~\cite{Poulin:2024ken}.
  
\end{itemize}

\section{Hubble tension and the DDR} \label{sec:results}

\subsection{$\Lambda$CDM background}
\label{sec:results_lcdm}
\begin{figure*}
      \includegraphics[width=\textwidth]{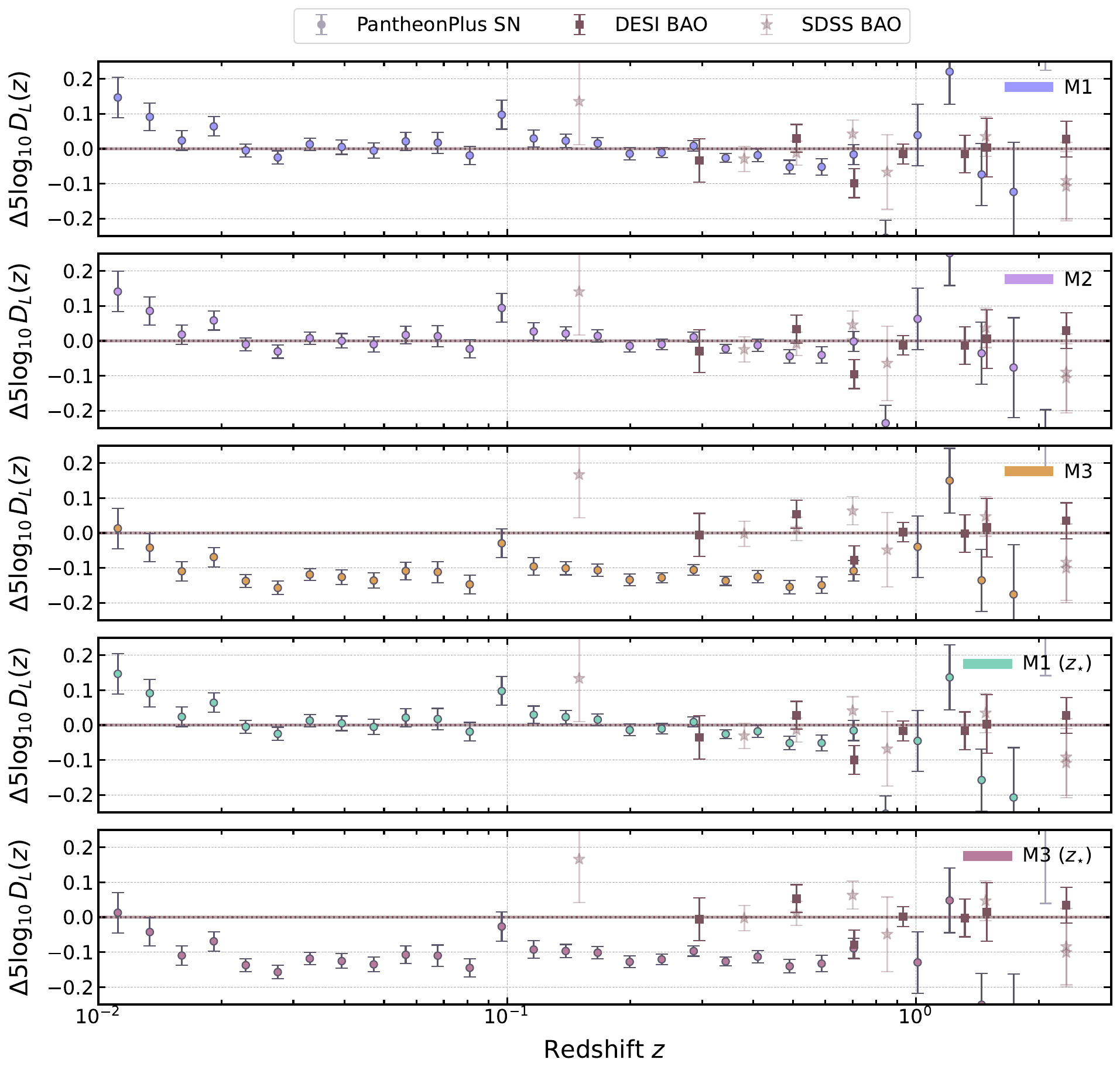}
  \caption{\label{fig:sn_bao_lcdm_ddr} Same as \cref{fig:sn_bao_lcdm_only} but for the different models of DDR breaking detailed in \cref{sec:method} under a $\Lambda$CDM background cosmology. The data points are normalised to the corresponding prediction with $\eta=1$ to highlight the effect of breaking the DDR in bringing the datasets together. We note that the observed offset is a combination of the inclusion of the DDR and different predictions for $M_B$, which illustrates the mismatch/compatibility between the two datasets.}
\end{figure*}

We now present our results examining the effect of DDR violation on the cosmic calibration, or Hubble tension. For the models with a transition redshift, we present and discuss the results obtained by fixing $z_* = 0.9$, which corresponds to the approximate median of the BAO redshift range, thereby splitting the data into two bins with similar statistical power. We briefly describe the effect of changing this transition redshift on our results in \cref{appendix:redshift}.

In \cref{fig:sn_bao_lcdm_only}, we have shown how in $\Lambda$CDM with no DDR violation
the various BAO estimates of $D_L(z)$ (in brown) are  systematically larger than the
SNIa estimates (in grey). In \cref{fig:sn_bao_lcdm_ddr}, we repeat the exercise for different models, plotting the residuals of the BAO and SNIa model with respect to the cosmology that best fits the combination of datasets (including the S$H_0$ES prior). We make use of the DDR violation to `move' the SNIa points, which are all anchored on the $M_B$ measured by S$H_0$ES. 
A model able to accommodate the S$H_0$ES measurement would thus have zero residuals. 
One can see that, depending on the functional form of $\eta$, the distance diagrams can be brought in good agreement.

\begin{figure*}
    \includegraphics[height=0.25\textwidth]{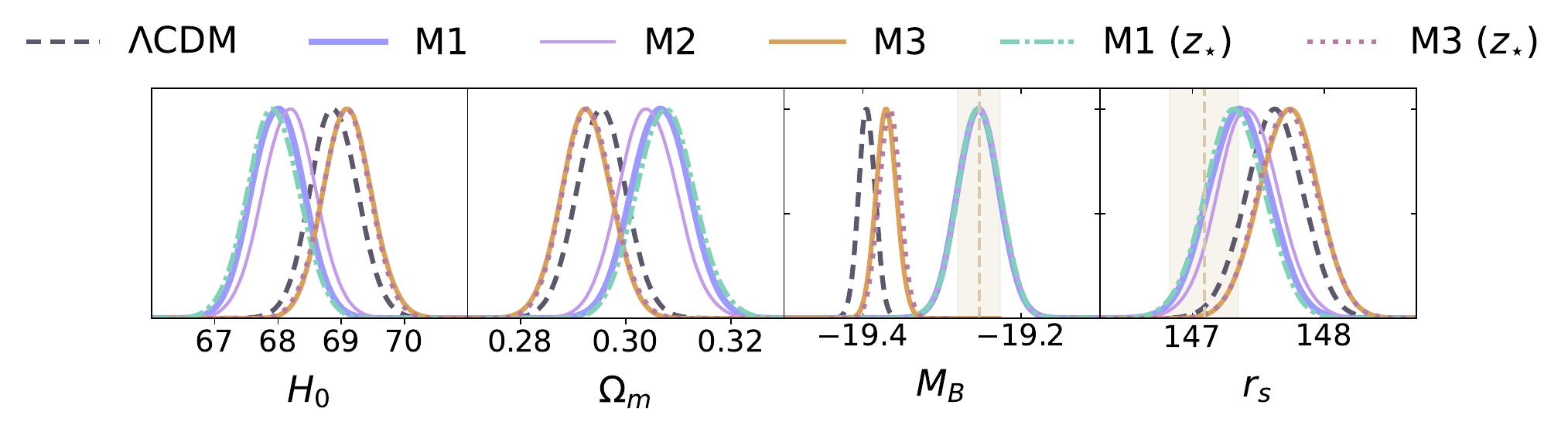}
  \caption{\label{fig:sn_bao} One-dimensional marginalised posterior distributions of the relevant cosmological parameters for the different DDR breaking models in a $\Lambda$CDM background using the combination of \textit{Planck} 2018 + DESI + PantheonPlus + S$H_0$ES prior data. The shaded bands in the $M_B$ and $r_s$ panels show the SNIa absolute magnitude used in the S$H_0$ES calibration and the sound horizon inferred from \textit{Planck}, respectively.}
\end{figure*}

In \cref{fig:sn_bao} we show the 1D marginalised posterior distributions of $\{H_0,\Omega_m, M_B, r_s\}$ for the five models \{M1, M2, M3, M1($z_{*}$), M3($z_{*}$)\} described in \cref{sec:param}, obtained in a $\Lambda$CDM background with our baseline dataset of \textit{Planck} 2018 + DESI + PantheonPlus + S$H_0$ES. The shaded bands in the $M_B$ and $r_s$ panels show the SNIa absolute magnitude used in the S$H_0$ES calibration and the sound horizon inferred from \textit{Planck}, respectively. From this figure, we can see that models M1 and M1($z_*$) both result in a value of $M_B$ in agreement with that of S$H_0$ES, whilst the value of $r_s$ inferred in these models agrees with that of \textit{Planck}. This demonstrates how the cosmic calibration tension can be interpreted as a violation of the DDR. 

In \cref{tab:all_models_LCDM_wprior_short} we list all the parameter constraints obtained in our models in the $\Lambda$CDM background, along with the Bayesian evidence $\mathcal{Z}_M$ normalised to $\Lambda$CDM with $\eta=1$ (no DDR break), $\log \mathcal{Z}_M/\mathcal{Z}_{\Lambda{\rm CDM}}$. Positive values indicate evidence for the DDR breaking model, whilst negative values indicate evidence for $\eta(z)=1$. The model with the largest number is the one favoured by the data we have used. To interpret the evidence ratios, we compare them to the modified Jeffreys scale~\citep{Jeffreys,Nesseris:2012cq}, whereby $X = \log \mathcal{Z}_{M}/\mathcal{Z}_{\Lambda{\rm CDM}} = [0, 1.1)$ indicates weak evidence for $M$, $X = [1.1, 3.0)$ indicates definite evidence for $M$, $X=[3.0, 5.0)$ is strong evidence and $X\ge5.0$ is very strong. 

The values listed in \cref{tab:all_models_LCDM_wprior_short} show that all DDR-violating models are preferred over $\Lambda$CDM under the datasets considered. The evidence is very strong for M1, M2 and M1($z_*$), definite for M3($z_*$) and weak for M3. Furthermore, the $\Delta \chi^2_{\rm min} = \chi^2_{{\rm min},\, \rm M} - \chi^2_{{\rm min},\, \Lambda{\rm CDM}}$ for each model is negative in each case, indicating that the DDR violating models provide a better fit to the data than $\Lambda$CDM. The parameter $\alpha_0$ is detected with a significance of $3-5\sigma$ for all models, with values that vary from $\alpha_0\sim-0.05$ (M3) to $\alpha_0\sim -0.075$ (M1), as expected from the relative difference in $H_0$ values between {\it Planck} $\Lambda$CDM and S$H_0$ES.
In addition, we note that there is some support in favour of a break in the DDR violation around $z\sim 1$, with $\Delta \alpha \equiv \alpha_0-\alpha_1 > 0$ at $\sim 2\sigma$ for both M1$(z_*)$ and M3$(z_*)$  (see App.~\ref{appendix:redshift} for details). This is in agreement with Ref.~\cite{Perivolaropoulos:2023iqj}, where evidence was found for two distinct values of the SNIa absolute magnitude $M_B$ taken at different redshift bins. However, we find that this is not favoured from an evidence standpoint compared to a constant offset.

To assess the amount of residual tension, we repeat the analysis above but without the S$H_0$ES prior included. We compute the Gaussian tension with the $M_B$ measurement from S$H_0$ES, 
\begin{equation}\label{eq:GT}
    {\rm GT}\equiv\bigg(\bar{M}_B-\bar{M}_B^{{\rm S}H_0{\rm ES}}\bigg)\bigg/\sqrt{\sigma^2+\sigma_{{\rm S}H_0{\rm ES}}^2} \mathcomma
\end{equation}
quantified as the difference in the mean value of the posterior of $M_B$ for the model being considered, $\bar{M}_B$, and the value measured by S$H_0$ES, $\bar{M}_B^{{\rm S}H_0{\rm ES}}$, divided by the respective quadrature sum of the standard deviations $\sigma^2$ and $\sigma_{{\rm S}H_0{\rm ES}}^2$, respectively. 
The results of this analysis are listed in \cref{tab:all_models_LCDM_noprior_short}.

As is well known, $\Lambda$CDM under \textit{Planck} is in tension with S$H_0$ES at more than $5\sigma$. This severe tension remains in M3 and M3($z_*$). However, the tension is resolved in M1, M2 and M1($z_*$). 
This confirms what is shown in \cref{fig:sn_bao_lcdm_ddr}, where the effect of the DDR breaking is to `move' the SNIa data points, such that they end up in agreement with the BAO data. 

From our results, we can conclude that a phenomenological violation of the DDR can indeed resolve the Hubble tension. However, the model with the largest evidence, M1 in a $\Lambda$CDM background, is compatible with a simple rescaling of the calibration of either SNIa or BAO (following \cref{eq:rs_rescale} or \cref{eq:Mb_rescale}).  This reinforces the interpretation of the Hubble tension as a `cosmic calibration tension'. 

\begin{table*}
\begin{center}
\renewcommand{\arraystretch}{1.5}
\resizebox{\textwidth}{!}{
\begin{tabular}{l c c c c c c c c c c c c c c c }
\hline
\multicolumn{7}{c}{$\Lambda$CDM + $\eta(z)$ for \textit{Planck} 2018 + DESI + PantheonPlus + S$H_0$ES prior} \\
\hline
\textbf{Parameter} & \textbf{ $\Lambda$CDM } & \textbf{ M1 } & \textbf{ M2 } & \textbf{ M3 } & \textbf{ M1 ($z_{*}=0.9$) } & \textbf{ M3 ($z_{*}=0.9$) } \\ 
\hline\hline
$ \alpha_0  $ & $ -- $ & $ -0.075\pm 0.012 $ & $ -0.070\pm 0.013 $ & $-0.049\pm 0.015$ & $ -0.076\pm 0.012 $ & $ -0.066\pm 0.017 $ \\ 
$ \alpha_1  $ & $ -- $ & $ -- $ & $ -0.014\pm 0.010 $ & $--$ & $ -0.039\pm 0.024 $ & $ 0.010\pm 0.026 $ \\ 
\hline
$M_B$ & $ -19.395\pm 0.011 $ & $ -19.254\pm 0.026 $ & $ -19.253\pm 0.026 $ & $ -19.370\pm 0.013$ & $ -19.253\pm 0.027 $ & $-19.367\pm 0.013$ \\ 
$ r_s  $ & $ 147.62\pm 0.22 $ & $ 147.35\pm 0.23 $ & $  147.42\pm 0.23 $ & $147.73\pm 0.22$ & $ 147.32\pm 0.23 $ & $  147.72\pm 0.22 $ \\ 

$ H_0  $ & $ 68.89\pm 0.38 $ & $ 68.00\pm 0.41  $ & $  68.18\pm 0.42 $ & $ 69.10\pm 0.38 $ & $  67.93\pm 0.41 $ & $  69.08\pm 0.38 $ \\ 
$ \Omega_m  $ & $ 0.2953\pm 0.0048 $ & $ 0.3066\pm 0.0054  $ & $  0.3042\pm 0.0055 $ & $0.2925\pm 0.0047$ & $  0.3076\pm 0.0055 $ & $ 0.2928\pm 0.0047 $ \\ 
\hline 
$\Delta \chi^2_{\text{min}}$ & $ -- $ & $ -32.84 $ & $ -34.73 $ & $ -10.27 $ & $ -35.40 $  & $ -17.25 $ \\
$\log \mathcal{Z}_M/\mathcal{Z}_{\Lambda{\rm CDM}}$ & $ 0 $ & $ 13.6 $ & $ 10.9 $ & $2.3$ & $ 12.3 $  & $3.3$ \\
\hline \hline
\end{tabular} }
\end{center}
\caption{Observational constraints at a $68 \%$ confidence level on the cosmological parameters for a $\Lambda$CDM cosmology with different models of DDR violation, inferred from analyses of the combination of \textit{Planck} 2018 data, DESI BAO and PantheonPlus SNIa calibrated with a S$H_0$ES prior.}
\label{tab:all_models_LCDM_wprior_short}
\end{table*}

\begin{table*}
\begin{center}
\renewcommand{\arraystretch}{1.5}
\resizebox{\textwidth}{!}{
\begin{tabular}{l c c c c c c c c c c c c c c c }
\hline
\multicolumn{7}{c}{$\Lambda$CDM + $\eta(z)$ for \textit{Planck} 2018 + DESI + PantheonPlus} \\
\hline
\textbf{Parameter} & \textbf{ $\Lambda$CDM } & \textbf{ M1 } & \textbf{ M2 } & \textbf{ M3 } & \textbf{ M1 ($z_{*}=0.9$) } & \textbf{ M3 ($z_{*}=0.9$) } \\ 
\hline\hline

$ \alpha_0  $ & $ -- $ & $ -0.088^{+0.096}_{-0.11} $ & $ -0.062^{+0.12}_{-0.090} $ & $-0.024\pm 0.016$ & $  -0.066^{+0.12}_{-0.086} $ & $ -0.040\pm 0.017 $ \\ 
$ \alpha_1  $ & $ -- $ & $ -- $ & $ -0.014\pm 0.010 $ & $--$ & $ -0.027^{+0.12}_{-0.094} $ & $ 0.026\pm 0.027 $ \\ 
\hline
$M_B$ & $ -19.422\pm 0.012 $ & $ -19.21\pm 0.23 $ & $ -19.26^{+0.19}_{-0.29} $ & $-19.407\pm 0.015$ & $ -19.27^{+0.19}_{-0.29} $ & $-19.402\pm 0.015$ \\ 
$ r_s  $ & $ 147.36\pm 0.23 $ & $ 147.35\pm 0.22 $ & $  147.43\pm 0.23 $ & $147.44\pm 0.23$ & $ 147.32\pm 0.23 $ & $  147.44\pm 0.23 $ \\ 
$ H_0  $ & $ 68.00\pm 0.40 $ & $ 68.00\pm 0.39  $ & $  68.17\pm 0.41 $ & $68.20\pm 0.42$ & $  67.91\pm 0.41 $ & $  68.22\pm 0.42 $ \\ 
$ \Omega_m  $ & $ 0.3065\pm 0.0054 $ & $ 0.3065\pm 0.0052  $ & $  0.3043\pm 0.0054 $ & $ 0.3039\pm 0.0055$ & $  0.3078\pm 0.0054 $ & $  0.3036\pm 0.0055 $ \\ 
\hline 
$\Delta \chi^2_{\text{min}}$ & $ -- $ & $ -0.33 $ & $ -2.44 $ & $ -2.51 $ & $ -3.34 $  & $ -7.36 $ \\
$\log \mathcal{Z}_M/\mathcal{Z}_{\Lambda{\rm CDM}}$ & $ 0 $ & $ -0.6 $ & $ -3.1 $ & $-2.1$ & $ -2 $  & $-2.1$ \\
GT w/ S$H_0$ES $M_B$ & $5.7\sigma $ & $ 0.2\sigma$ & $ 0.03\sigma $ & $5.0\sigma$ & $ 0.08\sigma $  & $4.8\sigma$ \\
\hline \hline
\end{tabular} }
\end{center}
\caption{Same as \cref{tab:all_models_LCDM_wprior_short} without the S$H_0$ES $M_B$ prior. We also report the Gaussian tension (GT) with the S$H_0$ES $M_B$ measurement defined in \eqref{eq:GT}.}
\label{tab:all_models_LCDM_noprior_short}
\end{table*}

\subsection{${w_0w_a}$CDM background}\label{sec:DDE}

The results described in the previous section were obtained in a $\Lambda$CDM background cosmology. We now investigate how applying a $w_0w_a$CDM background changes our constraints on the DDR and cosmological parameters and the corresponding evidence for DDR violation.

\begin{figure*}
    \includegraphics[height=0.25\textwidth]{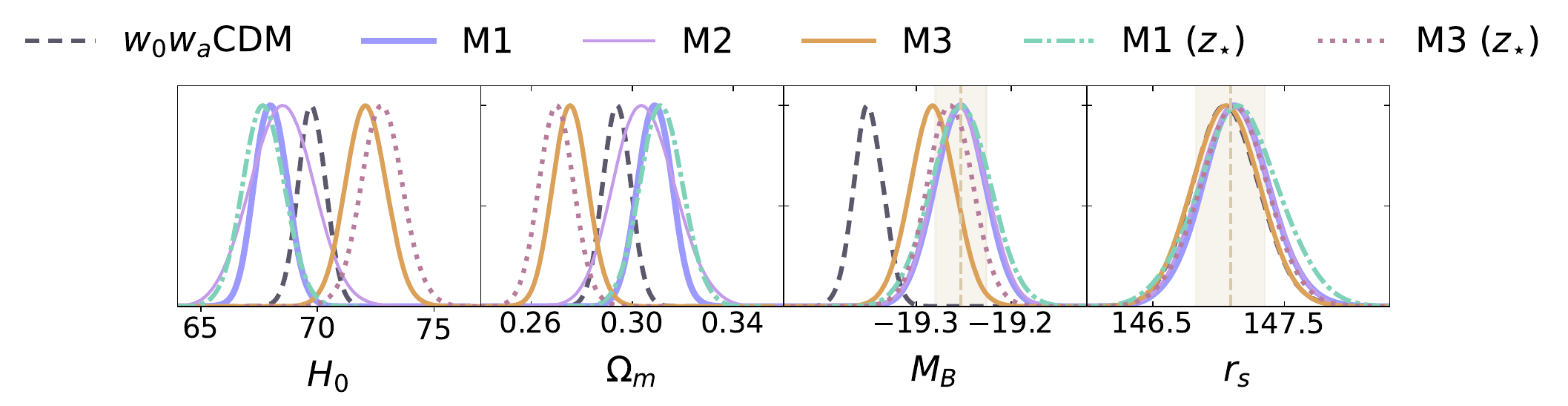}
  \caption{\label{fig:sn_bao_w0wa} One-dimensional marginalised posterior distributions of the relevant cosmological parameters for the different DDR breaking models in a  $w_0w_a$CDM background using the combination of  \textit{Planck} 2018 + DESI + PantheonPlus + S$H_0$ES prior data. The shaded bands in the $M_B$ and $r_s$ panels show the SNIa absolute magnitude used in the S$H_0$ES calibration and the sound horizon inferred from \textit{Planck}, respectively.}
\end{figure*}

\begin{figure*}
       \includegraphics[height=0.145\textwidth]{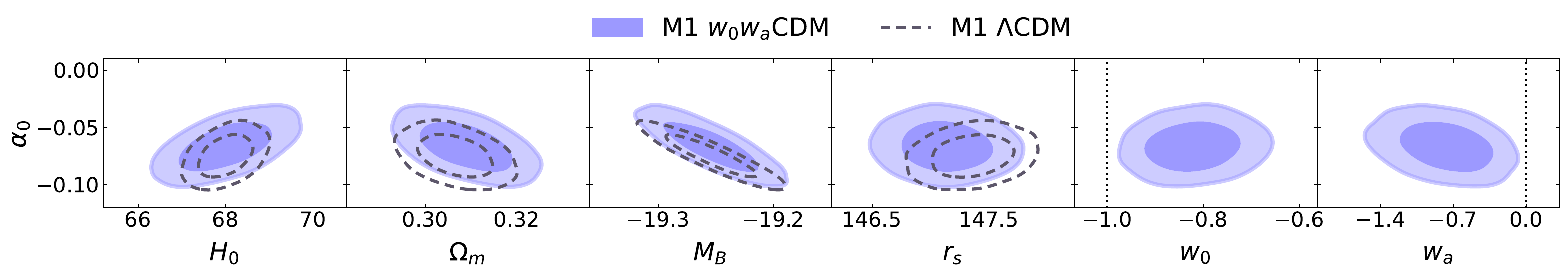}
      \includegraphics[height=0.145\textwidth]{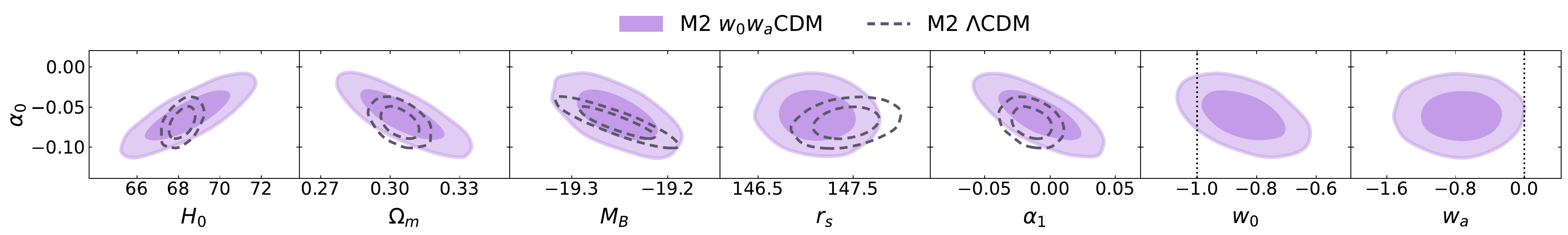}
      \includegraphics[height=0.145\textwidth]{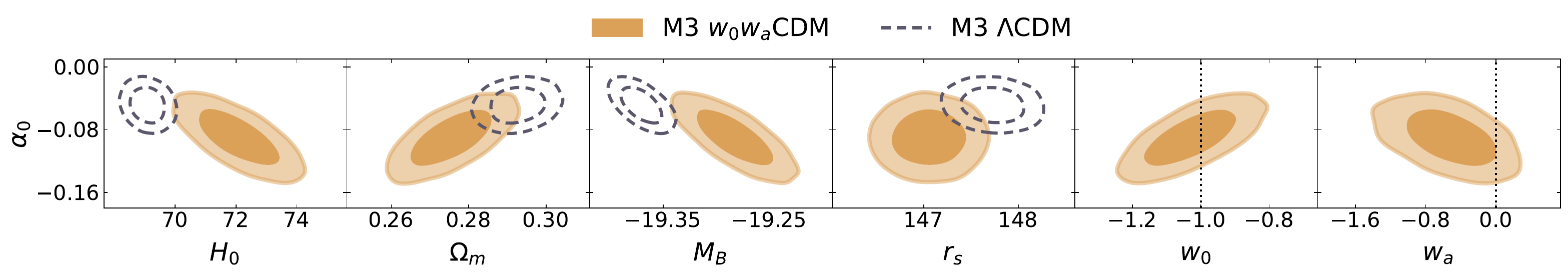}
      \includegraphics[height=0.145\textwidth]{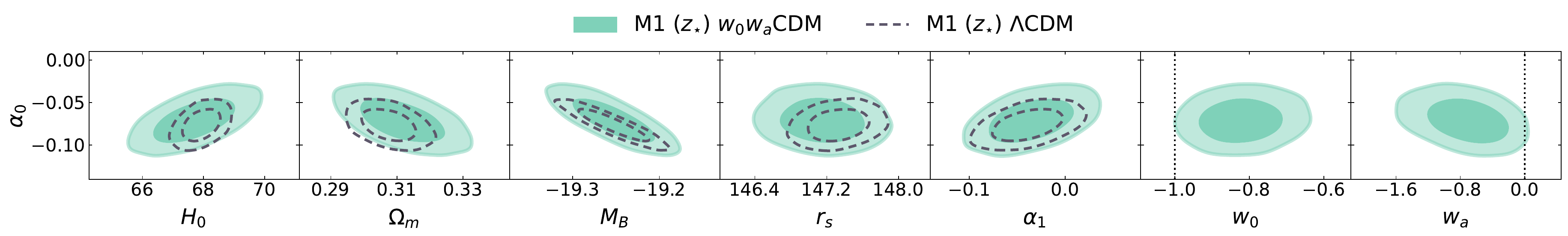}
      \includegraphics[height=0.145\textwidth]{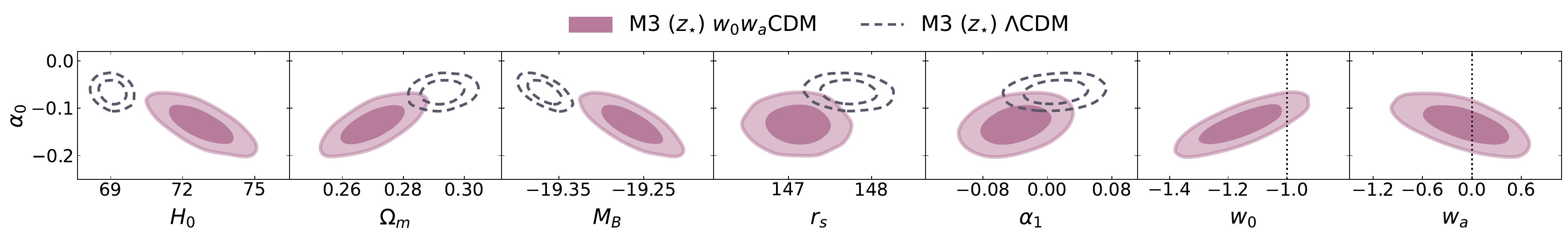}
  \caption{\label{fig:all_2d_fluid} Two-dimensional joint marginalised posterior distributions for the cosmological, DDR and fluid parameters for each model with $w_0w_a$CDM and $\Lambda$CDM background under  \textit{Planck} 2018 + DESI + PantheonPlus + S$H_0$ES prior data. The dashed vertical lines indicate the $\Lambda$CDM limit of the $w_0w_a$ parametrisation.}
\end{figure*}

\begin{figure*}
      \includegraphics[width=\textwidth]{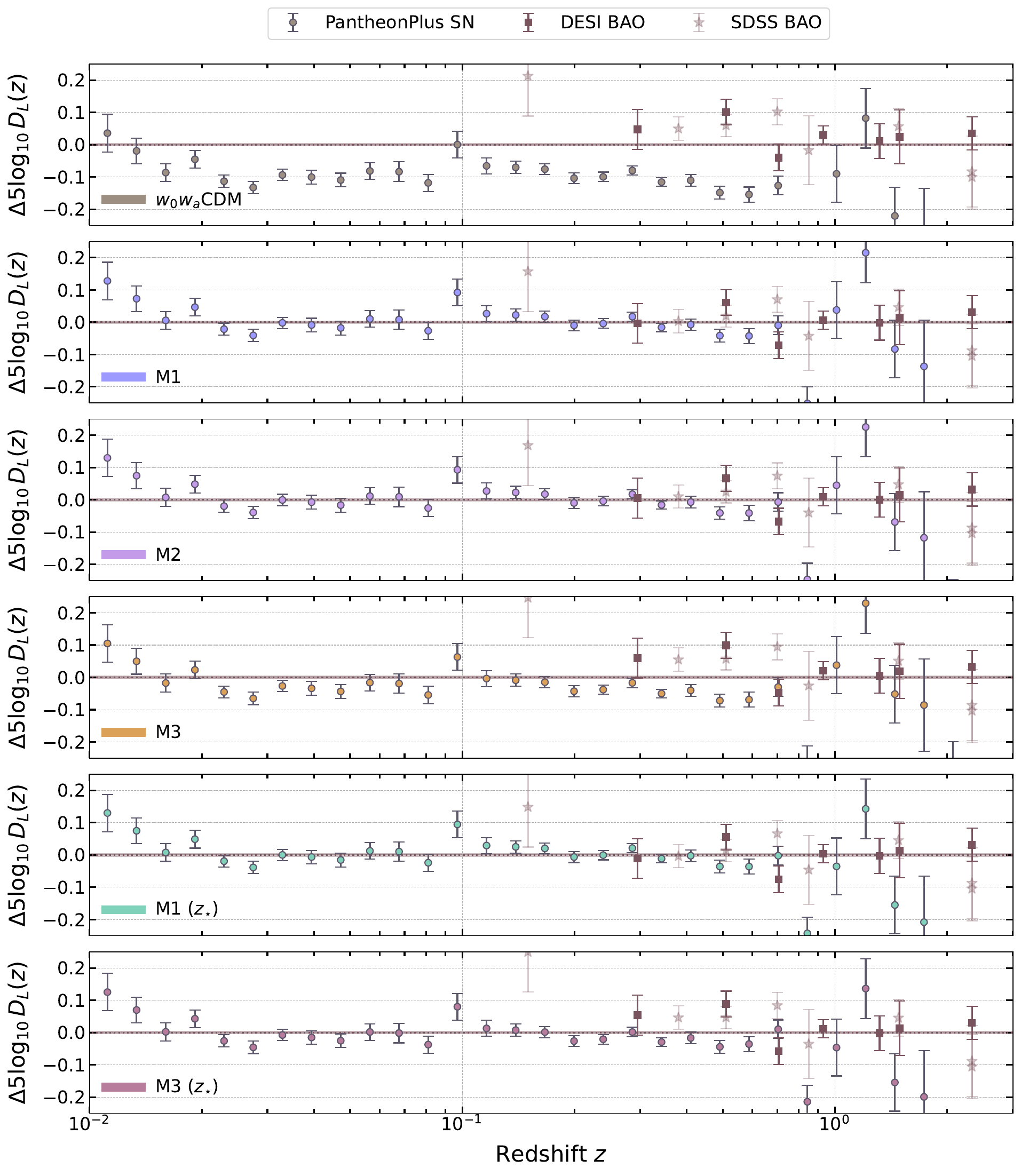}
  \caption{\label{fig:sn_bao_w0wa_panels} Same as \cref{fig:sn_bao_lcdm_only,fig:sn_bao_lcdm_ddr} but in the $w_0 w_a$ background cosmology.}
\end{figure*}

In \cref{fig:sn_bao_w0wa}, we show the same constraints as in \cref{fig:sn_bao} but obtained in a $w_0w_a$CDM cosmology.
The constraints on the parameters are listed in \cref{tab:all_models_w0wa_wprior_short}. As in the case of $\Lambda$CDM, we also ran our analysis without including the $M_B$ prior from S$H_0$ES, allowing us to quantify the remaining tension with that dataset. These results are listed in \cref{tab:all_models_w0wa_noprior_short}.\footnote{Comparing the $w_0w_a$ model to $\Lambda$CDM, we find that $\Lambda$CDM remains favoured, despite the improvement in $\chi^2$ with $w_0w_a$. This differs from what is reported in Ref.~\cite{DESI:2024mwx}, which found a similar $\Delta \chi^2$ but evidence favouring $w_0w_a$. These differences may come from differences in the prior on $w_0w_a$, as we do not impose $w_0+w_a<0$, slight differences in the dataset, or in the nested sampling algorithm itself.}
To investigate the role of the choice of background cosmology in the constraints, we compare in \cref{fig:all_2d_fluid} the 2D marginalised posteriors of $\alpha_0$ against those same parameters, both in $\Lambda$CDM  and in the $w_0w_a$ cosmology. We also include panels showing $\alpha_0$-vs-$\{w_0,w_a\}$ for all models, and $\alpha_0$-vs-$\alpha_1$ when applicable. We recall that values of $\alpha_{0,1}\neq0$ suggest deviations from the standard DDR. As expected, the inclusion of the two dynamical dark energy parameters weakens the constraints on the DDR parameters $\alpha_{0,1}$. In contrast to the results in a $\Lambda$CDM background, the flexibility introduced to the background history implies that all our models of DDR violation now produce a sound horizon size that agrees with that of \textit{Planck} and a SNIa absolute magnitude that agrees with that of S$H_0$ES, reflecting the resolution of the tension. This is particularly evident for the M3 and M$3(z_*)$ models in \cref{fig:all_2d_fluid}. Moreover, we find evidence for evolving dark energy in all models at more than $1\sigma$ except for M$3(z_*)$, which shows a preference for a constant phantom equation of state parameter. Consequently, the distances are now in good agreement, as shown in \cref{fig:sn_bao_w0wa_panels}.
This is further supported by the GT metric, which yields a $2.4\sigma$ residual tension for the M3 model at worst and 1.1$\sigma$ or better (\textit{i.e.}, perfect agreement) for all other models.

\begin{figure*}
      \includegraphics[width=\textwidth]{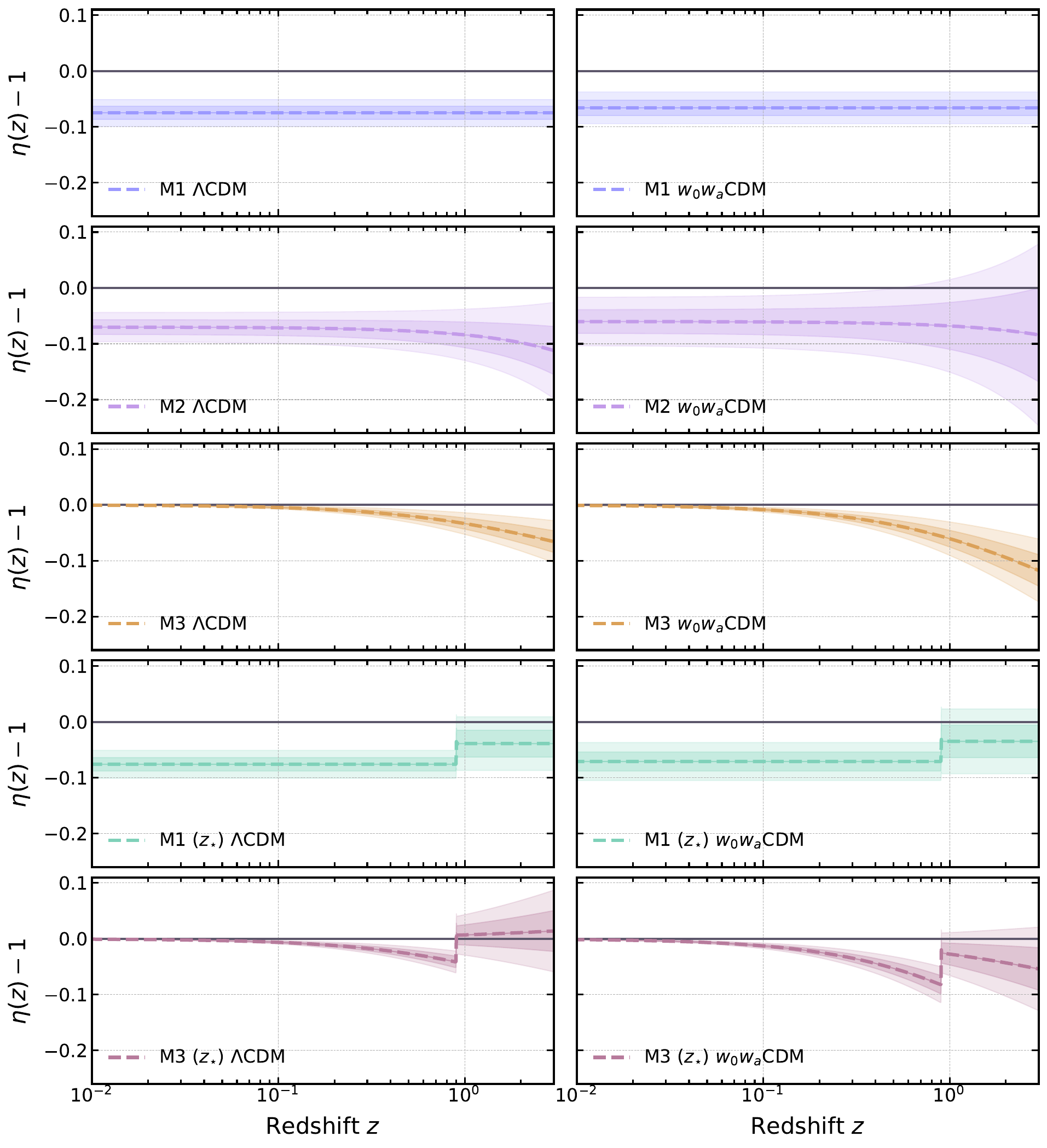}
  \caption{\label{fig:sn_bao_lcdm_w0wa_etas} Redshift evolution of each of the functions $\eta(z)$ as defined in \cref{eq:eta_ddr} for the different DDR-breaking models listed in \cref{sec:param}. The curves are estimated from the best-fit parameters for each model detailed in \cref{tab:all_models_LCDM_wprior_short,tab:all_models_w0wa_wprior_short} under a $\Lambda$CDM (left panels) and $w_0w_a$ (right panel) background cosmology for the  \textit{Planck} 2018 + DESI + PantheonPlus + S$H_0$ES prior data. The coloured bands refer to the 68\% and 95\% confidence regions.}
\end{figure*}

In \cref{fig:sn_bao_lcdm_w0wa_etas}, we show the function $\eta(z)-1$ for each of our five models in a $\Lambda$CDM background (left column), and a $w_0w_a$CDM background (right column), for the data combination that includes the S$H_0$ES prior. The results are consistent in both cosmological backgrounds for a constant violation of the DDR, M1. The linear violation of the DDR, M2, is the same in both cosmologies at low redshift, but in the $w_0w_a$CDM background, the function becomes consistent with $\eta(z)=1$ \textit{i.e.} no DDR violation at $z\gtrsim 2$. For the power-law DDR violation, M3, both cosmologies favour similar functional forms for $\eta(z)$, with the $w_0w_a$CDM background showing a more substantial deviation from $\eta(z)=1$ at $z > 1$. 

For a constant DDR violation that changes above some transition redshift, M1($z_*$), the results in the two background cosmologies are again consistent with each other. The functions of the power-law with transition, M3($z_*$), however, are different. In the $\Lambda$CDM background, there is a hint for $\eta(z)-1 > 0$ above $z=1$, which sharply transitions to a negative value before once again becoming consistent with no DDR breaking at late times. However, in the  $w_0w_a$CDM background, $\eta(z)-1$ is always negative but growing. It also becomes consistent with no DDR violation at $z\rightarrow 0$.

Results for the M3$(z_*)$ model suggest that a phantom dark energy with $w\sim -1.155$ and a deviation in the DDR affecting the data at $z \lesssim 0.9$ with $\alpha_0\sim -0.134$ (and $\alpha_1\sim 0$ at $1.5\sigma$)  provide a good fit to the data, with $w_a\sim 0$. Given that the parameters $\alpha_1$ and $w_a$ are compatible with zero at $\sim1\sigma$, it is instructive to run one additional analysis enforcing $\alpha_1 = w_a = 0$ in the M$3(z_*)$ scenario. In this case, we find $\log \mathcal{Z}_M/\mathcal{Z}_{\Lambda{\rm CDM}} = 15.4$, which performs {\it better than all models}, including a definite preference over M1 for both a $\Lambda$CDM background or $w_0w_a$CDM background. 

This is reminiscent of the model discussed in Ref.~\cite{Tutusaus:2023cms}, where an exotic dark energy model, together with a violation in the DDR (therein interpreted as a systematic in the high-$z$ SNIa measurements), was found to explain the Hubble tension. Future data, able to break the degeneracy between DDR violation and the background expansion (e.g. longitudinal BAO data at all $z$, or extending the BAO measurements to lower $z$), will be crucial to disentangling the M1 and M3$(z_*)$ scenarios.

\begin{table*}
\begin{center}
\renewcommand{\arraystretch}{1.5}
\resizebox{\textwidth}{!}{
\begin{tabular}{l c c c c c c c c c c c c c c c }
\hline
\multicolumn{7}{c}{$w_0w_a$ + $\eta(z)$ for \textit{Planck} 2018 + DESI + PantheonPlus + S$H_0$ES prior} \\
\hline
\textbf{Parameter} & \textbf{ $w_0w_a$ } & \textbf{ M1 } & \textbf{ M2 } & \textbf{ M3 } & \textbf{ M1 ($z_{*}=0.9$) } & \textbf{ M3 ($z_{*}=0.9$) } \\ 
\hline\hline

$ \alpha_0  $ & $ -- $ & $ -0.066\pm 0.014 $ & $ -0.060\pm 0.021 $ & $-0.090\pm 0.023$ & $ -0.071\pm 0.017 $ & $ -0.134\pm 0.028\ [-0.128 \pm 0.021] $ \\ 
$ \alpha_1  $ & $ -- $ & $ -- $ & $ -0.008\pm 0.021 $ & $--$ & $ -0.035\pm 0.029 $ & $ -0.040\pm 0.029\ [--] $ \\ 
\hline 
$ w_0  $ & $ -0.784\pm 0.067 $ & $ -0.820\pm 0.065 $ & $ -0.846\pm 0.091 $ & $-1.029\pm 0.089$ & $ -0.821\pm 0.075 $ & $ -1.155\pm 0.095\  [-1.158 \pm 0.033] $ \\ 
$ w_a  $ & $ -1.20^{+0.34}_{-0.30} $ & $ -0.78^{+0.32}_{-0.27} $ & $ -0.74^{+0.33}_{-0.29} $ & $ -0.52^{+0.36}_{-0.32}$ & $ -0.72^{+0.36}_{-0.30} $ & $ -0.09^{+0.37}_{-0.32}\ [--] $ \\ 
\hline 
$M_B$ & $ -19.350\pm 0.016 $ & $ -19.254\pm 0.027 $ & $ -19.254\pm 0.028 $ & $-19.283\pm 0.024$ & $ -19.253\pm 0.032 $ & $-19.264\pm 0.024$ \\ 
$ r_s  $ & $ 147.06\pm 0.25 $ & $ 147.13\pm 0.26 $ & $  147.12\pm 0.26 $ & $147.06\pm 0.26$ & $ 147.15\pm 0.31 $ & $  147.11\pm 0.26 $ \\ 
$ H_0  $ & $ 69.77\pm 0.60 $ & $ 68.00\pm 0.71  $ & $  68.5\pm 1.3  $ & $72.08\pm 0.88$ & $  67.70\pm 0.89 $ & $  72.77\pm 0.91 $ \\ 
$ \Omega_m  $ & $ 0.2939\pm 0.0055 $ & $ 0.3089\pm 0.0067  $ & $  0.305^{+0.011}_{-0.013} $ & $0.2756\pm 0.0070$ & $  0.3116\pm 0.0085 $ & $  0.2700\pm 0.0070 $ \\ 
\hline 
$\Delta \chi^2_{\text{min}}$ & $ -20.37 $ & $ -40.89 $ & $ -40.16 $ & $ -34.82 $ & $ -42.75 $  & $ -43.20\ [-41.86 ] $ \\
$\log \mathcal{Z}_M/\mathcal{Z}_{\Lambda{\rm CDM}}$ & $ 5.6 $ & $ 12.7 $ & $ 10.4$ & $10.5$ & $ 11.5 $  & $12.4\ [15.4]$ \\
\hline \hline
\end{tabular} }
\end{center}
\caption{ Same as \cref{tab:all_models_LCDM_wprior_short}, now in the $w_0w_a$CDM cosmology and with the S$H_0$ES $M_B$ prior. In square brackets, we list the values for M3 ($z_{*}=0.9$) with $\alpha_1=w_a=0$ (the corresponding cosmological parameters are consistent with the full case).}
\label{tab:all_models_w0wa_wprior_short}
\end{table*}

\begin{table*}
\begin{center}
\renewcommand{\arraystretch}{1.5}
\resizebox{\textwidth}{!}{
\begin{tabular}{l c c c c c c c c c c c c c c c }
\hline
\multicolumn{7}{c}{$w_0w_a$ + $\eta(z)$ for \textit{Planck} 2018 + DESI + PantheonPlus} \\
\hline
\textbf{Parameter} & \textbf{ $w_0w_a$ } & \textbf{ M1 } & \textbf{ M2 } & \textbf{ M3 } & \textbf{ M1 ($z_{*}=0.9$) } & \textbf{ M3 ($z_{*}=0.9$) } \\ 
\hline\hline
$ \alpha_0  $ & $ -- $ & $ -0.072\pm 0.096 $ & $ -0.057^{+0.12}_{-0.096} $ & $-0.012\pm 0.040$ & $ -0.065^{+0.12}_{-0.093} $ & $  -0.085\pm 0.054 $ \\ 
$ \alpha_1  $ & $ -- $ & $ -- $ & $ -0.008\pm 0.021 $ & $--$ & $ -0.03^{+0.12}_{-0.10} $ & $ -0.011\pm 0.040 $ \\ 
\hline 
$ w_0  $ & $ -0.822\pm 0.065 $ & $ -0.821\pm 0.063 $ & $ -0.846\pm 0.092 $ & $-0.85\pm 0.12$ & $ -0.825\pm 0.061 $ & $ -1.04\pm 0.15 $ \\ 
$ w_a  $ & $ -0.77\pm 0.30 $ & $ -0.77^{+0.31}_{-0.27} $ & $ -0.74^{+0.33}_{-0.29} $ & $-0.72^{+0.35}_{-0.32}$ & $ -0.69^{+0.29}_{-0.25} $ & $ -0.28\pm 0.39 $ \\ 
\hline 
$M_B$ & $ -19.404\pm 0.020 $ & $ -19.23^{+0.21}_{-0.27} $ & $ -19.25^{+0.20}_{-0.29} $ & $-19.390\pm 0.049$ & $ -19.26^{+0.20}_{-0.29} $ & $-19.323\pm 0.060$ \\ 
$ r_s  $ & $ 147.13\pm 0.26 $ & $ 147.13\pm 0.26 $ & $  147.12\pm 0.26 $ & $147.12\pm 0.26$ & $ 147.16\pm 0.26 $ & $  147.13\pm 0.26 $ \\ 
$ H_0  $ & $ 67.99\pm 0.71 $ & $ 68.00\pm 0.70  $ & $  68.4\pm 1.3 $ & $68.5\pm 1.7$ & $  67.65\pm 0.74 $ & $  70.8\pm 2.1 $ \\ 
$ \Omega_m  $ & $ 0.3091\pm 0.0068 $ & $ 0.3090\pm 0.0067  $ & $  0.305^{+0.011}_{-0.013} $ & $0.305\pm 0.015$ & $  0.3118\pm 0.0071 $ & $  0.286\pm 0.017 $ \\ 
\hline 
$\Delta \chi^2_{\text{min}}$ & $ -8.05 $ & $ -7.95 $ & $ -6.66 $ & $ -8.16 $ & $ -10.50 $  & $ -11.06 $ \\
$\log \mathcal{Z}_M/\mathcal{Z}_{\Lambda{\rm CDM}}$ & $ -1.2 $ & $ -1.4 $ & $ -6.1 $ & $-2.7$ & $ -3 $  & $-2.8$ \\
GT w/ S$H_0$ES $M_B$& $4.4\sigma$ & $0.08\sigma$ & $0.01\sigma$ & $2.4\sigma$ & $0.03\sigma$  & $ 1.1\sigma$ \\
\hline \hline
\end{tabular} }
\end{center}
\caption{ Same as \cref{tab:all_models_w0wa_wprior_short} without the S$H_0$ES $M_B$ prior. We also report the Gaussian tension (GT) with the S$H_0$ES $M_B$ measurement defined in \eqref{eq:GT}.}
\label{tab:all_models_w0wa_noprior_short}
\end{table*}

\section{Discussion} \label{sec:discussion}

\begin{figure*}
      \includegraphics[width=0.46\textwidth]{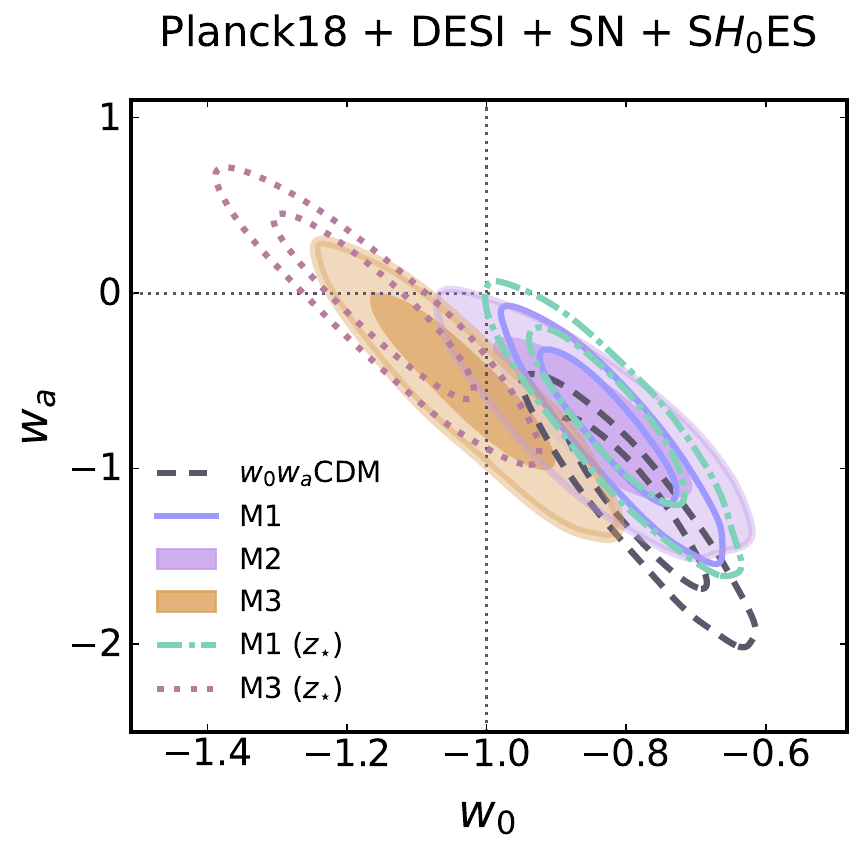} \hfill
      \includegraphics[width=0.46\textwidth]{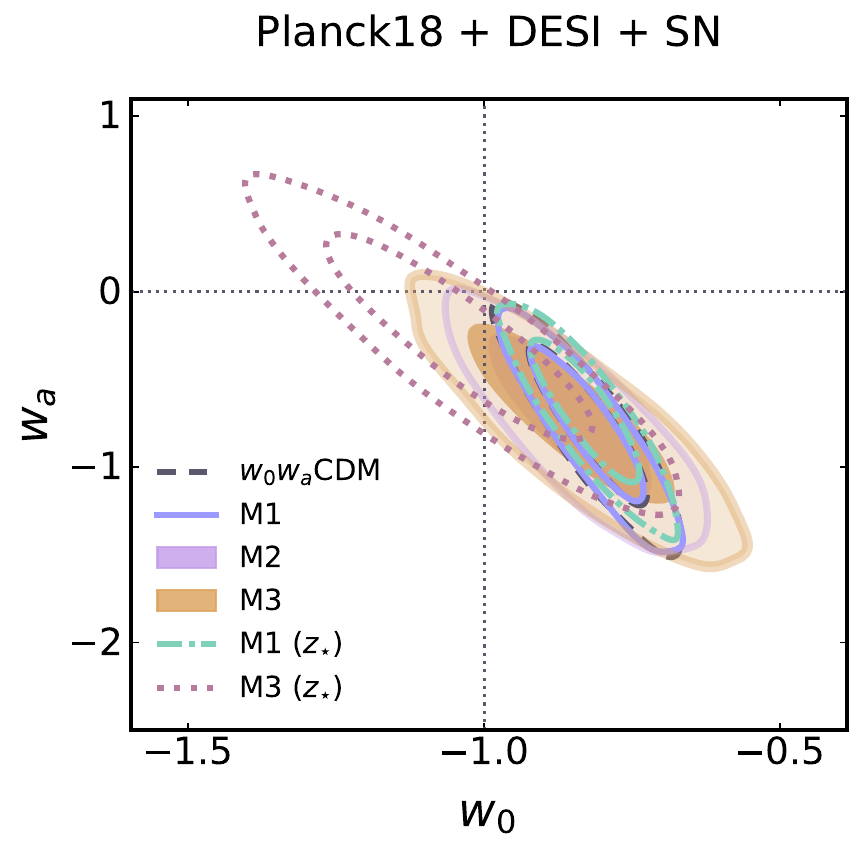}
  \caption{\label{fig:w0wa} Two-dimensional joint marginalised posterior distribution of the fluid parameters $w_0$ and $w_a$ for the different DDR breaking models under  \textit{Planck} 2018 + DESI + PantheonPlus data. The left panel includes a S$H_0$ES prior, while the right one does not.}
 \end{figure*}

\subsection{Degeneracy between DDR and dynamical dark energy}
Our results indicate that there is some degeneracy between a violation of the DDR and a change in the background cosmology that affects the expansion history $H(z)/H_0$. This may appear surprising as, in principle, a violation of the DDR ($\eta(z)\neq 1$ in \cref{eq:eta_ddr}) has different phenomenological effects than a deviation from a flat $\Lambda$CDM background. On the one hand, the background cosmology affects the luminosity and angular diameter distances in the same way and thus `moves' SNIa and transverse BAO data at the same redshift in the same way. On the other hand, a break in the DDR can help `move' SNIa and transverse BAO points at the same redshift relative to each other. Hence, it is not evident that a deviation from the $\Lambda$CDM background cosmology and a DDR violation could have a degenerate effect.

However, there are several reasons why degeneracy occurs. First, as BAO data can measure $D_H$ (see \cref{sec:ddrtheory}), which is inversely proportional to $H(z)$ rather than an integral of $H(z)$, a change in $H(z)$ can move the longitudinal BAO relative to the SNIa data. Second, and more importantly,  SNIa and BAO do not perfectly overlap in redshift: there are many more SNIa data at $0.01\lesssim z\lesssim 0.3$. As a result, changes in the DDR at such redshifts can arbitrarily move the SNIa points without affecting the BAO and thus mimic the effect of a change in $H(z)$. This suggests that the preference for evolving dark energy observed when combining DESI and SNIa data may be altered (or even removed) when simultaneously considering a model with DDR violation.

In \cref{fig:w0wa}, we show the 2D marginalised joint posterior distribution of $\{w_0,w_a\}$ in all five models compared with the fiducial case $\eta=1$, when including the S$H_0$ES prior on the left, and without it on the right. We can see that DDR violations can indeed significantly affect this joint posterior and the corresponding evidence for dynamical dark energy (\textit{i.e.}, $w_0\neq -1,w_a\neq 0$).  
Most importantly, in models M3$(z_*)$, the preference for dynamical dark energy is completely gone when the S$H_0$ES prior is left out of the analysis, with the $\Lambda$CDM limit recovered at $\sim 1\sigma$. 
This suggests that, with the data considered here, the preference for dynamical dark energy can be equally interpreted as a hint for a break in the DDR, with a preference for a deviation occurring at $z<z_*$ (\textit{i.e.}, $\alpha_0\neq0$, $\alpha_1\sim 0$). 
However, we note that results from DESI DR2 (which were released during the completion of this work) suggest a 3$\sigma$ preference for dynamical dark energy solely from the combination of CMB and BAO (\textit{i.e.} without SNIa) \cite{DESI:2025zgx}. Therefore, we anticipate that an updated analysis with DESI DR2 data would favour dynamical dark energy over a DDR violation, though the CMB+BAO result is still at fairly weak statistical significance. We will address this in a future work.

\subsection{A DDR deviation or a change in calibration?}
The two models favoured in a Bayesian sense are M1 in a flat $\Lambda$CDM background and M3($z_*$) in the $w_0w_a$CDM background.
These correspond to two very different scenarios with distinct observational consequences.  

Model M1 suggests that the SNIa magnitude measured by S$H_0$ES in the second rung of the distance ladder (\textit{i.e.} Cepheid $\rightarrow$ SNIa) may differ from that of high-$z$ SNIa, by an amount given by \cref{eq:Mb_rescale}: $\Delta M \sim -0.17$ Mag. This matches the difference in magnitude inferred from the inverse calibration under the $\Lambda$CDM fit to {\it Planck} and BAO ($M_B^{\Lambda{\rm CDM}} \sim -19.42$) compared to that measured by S$H_0$ES ($M_B^{{\rm S}H_0{\rm ES}} \sim -19.25$). Consequently, the value of $H_0$ is in agreement with {\it Planck} $\Lambda$CDM, with $H_0 \sim 68\pm0.4$ km/s/Mpc.

There are two ways of interpreting the result of M1. Firstly, if the `true' magnitude is that inferred from \textit{Planck} within $\Lambda$CDM, it suggests that the objects involved in the first or the second rung of the ladder are affected by some unknown physics that make them appear dimmer (\textit{e.g.} \cite{Rigault:2013gux,Desmond:2019ygn}). Currently, calibration of SNIa with Tip of the Red Giant Branch stars appears to yield results in agreement with Cepheids \cite{Riess:2021jrx, Riess:2024vfa}. Although some measurements appear to yield slightly different values of the SNIa magnitude \cite{Freedman:2024eph}, sample fluctuation and selection biases have been suggested as a simple explanation of the differences \cite{Riess:2024vfa}. There is also currently no robust confirmation of different SNIa properties between those observed by S$H_0$ES and those from larger SNIa catalogues like Pantheon+ \cite{Brout:2023wol}. Therefore, this explanation is currently disfavoured, though the best way to rule it out would be an independent measurement of $H_0$. 

Alternatively, if the `true' magnitude is the one measured by S$H_0$ES, the model M1 studied here can be viewed as phenomenological parameterisations of a mechanism that would augment the flux of the high-$z$ SNIa on their way to us. Indeed, if the flux we measure is increased by some mechanism when interpreted under the {\it Planck} $\Lambda$CDM cosmology that fixes them at a larger distance, they would appear intrinsically brighter (with magnitude $M_B^{\Lambda{\rm CDM}}$ rather than $M_B^{{\rm S}H_0{\rm ES}}$) than they really are. 

For this scenario to work, we would need a mechanism that affects all SNIa apart from the S$H_0$ES dataset of local SNIa; for example,  extra energy may be emitted in the form of invisible particles, such as axions, that escape in our local vicinity, but may later on, during propagation over large distances, convert into photons. However, this does not agree with our finding that the data favours a redshift-independent DDR violation, M1. There are also strong constraints on these scenarios from SNIa observations \cite{Crnogorcevic:2021wyj,Calore:2021hhn}.

Alternatively, the M3($z_*) + w_0w_a$ cosmology, suggests that, on top of a phantom dark energy component ($w_0\sim-1.155,w_a\sim 0$), some mechanism exists to alter the magnitude of the SNIa such that the DDR scales as $D_L = (1+z)^{2+\alpha} D_A(z)$, with $\alpha\sim -0.134$ at $z\lesssim 1$ and $\alpha\sim 0$ at higher $z$.
In this cosmology, the value of $H_0$ is large, $H_0 = 72.77\pm0.91$ km/s/Mpc in agreement with the fiducial S$H_0$ES analysis. 

These various scenarios may be disentangled from one another with higher accuracy data. Firstly, independent $H_0$ measurements confirming a high value will firmly establish that the new physics mechanisms cannot solely affect the SNIa magnitude inferred in the second rung of the distance ladder. This would rule out any sort of systematic error or new physics that solely affects the SNIa distance ladder. Currently, alternative methods, such as strong lensing time delays \cite{Wong:2019kwg} and results from the Megamaser Cosmology Project \cite{Pesce:2020xfe},  seem to point towards a larger value of $H_0$, although this is debated \cite{Birrer:2020tax,Vagnozzi:2023nrq}. Secondly, whilst the toy models we have used do not alter the CMB by construction, realistic models could leave imprints in, for instance, the redshift evolution of the CMB temperature \cite{Euclid:2020ojp} which would be different for different DDR violation models. Finally, given the very different background history, as one model invokes a cosmological constant whilst the other suggests a phantom dark energy component, new BAO and SNIa measurements at low-$z$ could help distinguish the two.

These scenarios also differ from `early-universe' models in which the sound horizon is altered. Those typically lead to high $H_0$ values, but also very different cosmological parameters \cite{Poulin:2024ken}, namely, a larger value of the matter density. They also lead to specific signatures in the CMB. Hence, intermediate to high-$\ell$ CMB measurements at higher accuracy \cite{Smith:2025zsg} and alternative methods measuring the values of $H_0$ and related parameters (\textit{e.g.} the age of the Universe, matter density) can be used to differentiate between the DDR violations we study here, and solutions that alter the sound horizon. 

\section{Conclusions} \label{sec:conclusions}

In this work, we have investigated whether a violation of the distance duality relation can resolve the Hubble tension. We have used a combination of data that includes \textit{Planck} CMB, DESI BAO and Pantheon+ supernovae calibrated with S$H_0$ES  to place constraints on five phenomenological parameterisations of the violation of the distance duality relation. We conducted our analysis in both a $\Lambda$CDM and $w_0w_a$CDM background to assess how the change in cosmology affects the constraints on the DDR violation models. 

We found that all the models we considered are preferred over the standard $\Lambda$CDM cosmology with no DDR violation, although not all can resolve the tension with S$H_0$ES. 
The data currently favours two possibilities: 
a constant violation of the DDR (equivalent to a calibration shift), $D_L(z)/D_A(z)\simeq 0.925(1+z)^2$; or a change in the power-law redshift-dependence of the DDR, restricted to $z\lesssim 1$, $D_L(z)/D_A(z)\simeq(1+z)^{1.866}$, together with a phantom dark energy equation of state $w\sim -1.155$ (similar to Ref.~\cite{Tutusaus:2023cms}). The data slightly favour the latter scenario over the former.

These scenarios lead to different values for $H_0$ and expansion histories, allowing us to disentangle models with future independent and precise measurements of $H_0$.  Furthermore, models invoking DDR violation can also be disentangled from  `early-universe' models that alter the sound horizon, as they lead to different cosmological parameters ($H_0$ and $\Omega_M$ in particular).

Our approach is phenomenological, and we leave the development of a physical model of DDR breaking that would produce such effects for future work. We anticipate that further constraints would apply to realistic scenarios, but our work provides motivation for model builders and a novel approach to resolving the `Hubble tension'.\\

\begin{acknowledgments}
We thank Chiara de Leo, Matteo Martinelli, Alessandro Melchiorri and Théo Simon for useful discussions.
EMT, TM, AP and VP are supported by funding from the European Research Council (ERC) under the European Union’s HORIZON-ERC-2022 (grant agreement no. 101076865). WG acknowledges support from the Lancaster–Sheffield Consortium for Fundamental Physics through the Science and Technology Facilities Council (STFC) grant ST/X000621/1. NBH is supported by a postdoctoral position funded by IN2P3. We gratefully acknowledge support from the CNRS/IN2P3 Computing Center (Lyon - France) for providing computing and data-processing resources needed for this work.
\end{acknowledgments}

\appendix

\section{Changing the transition redshift}\label{appendix:redshift}

\begin{figure*}
      \includegraphics[width=0.48\textwidth]{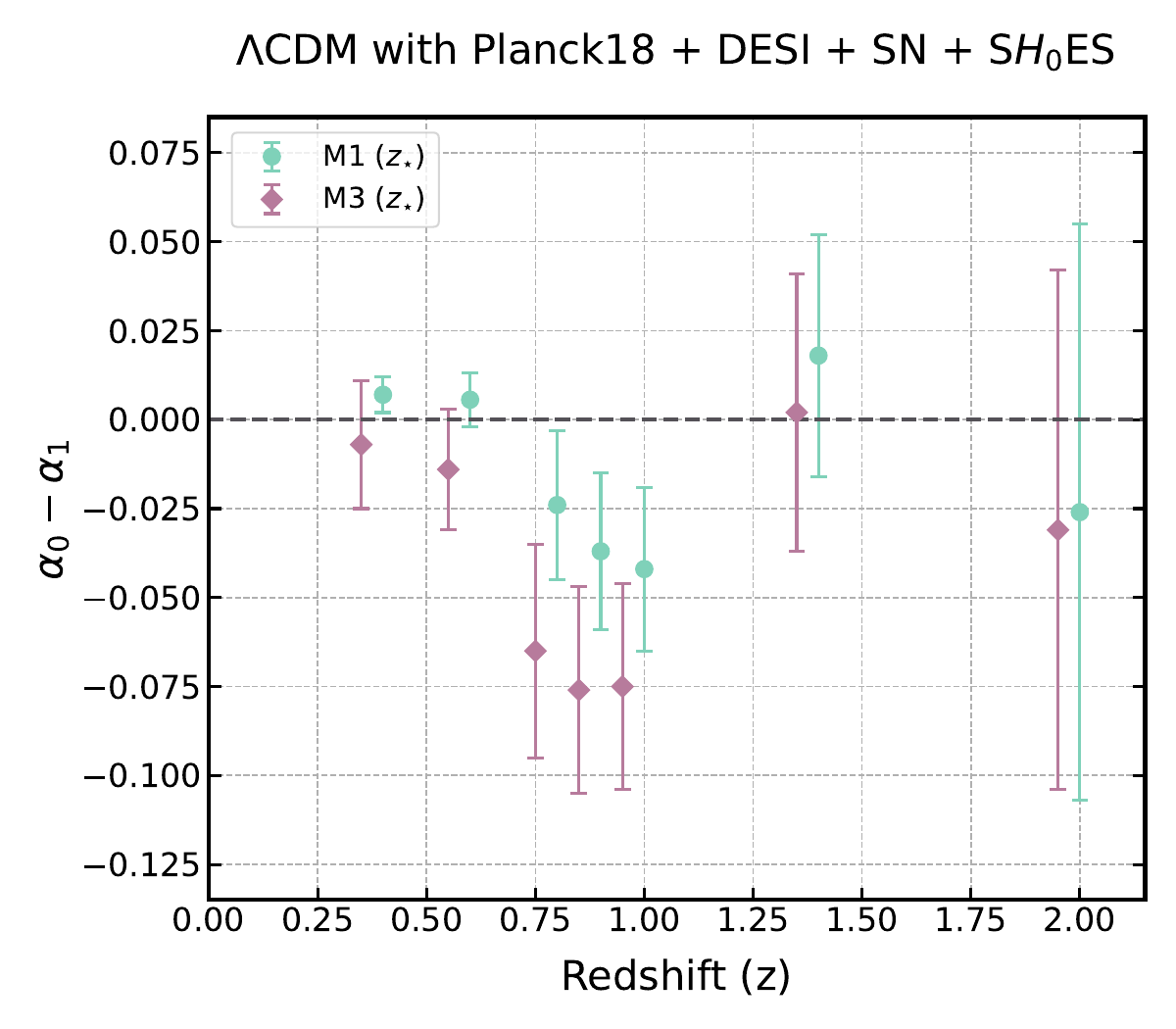} \hfill
      \includegraphics[width=0.48\textwidth]{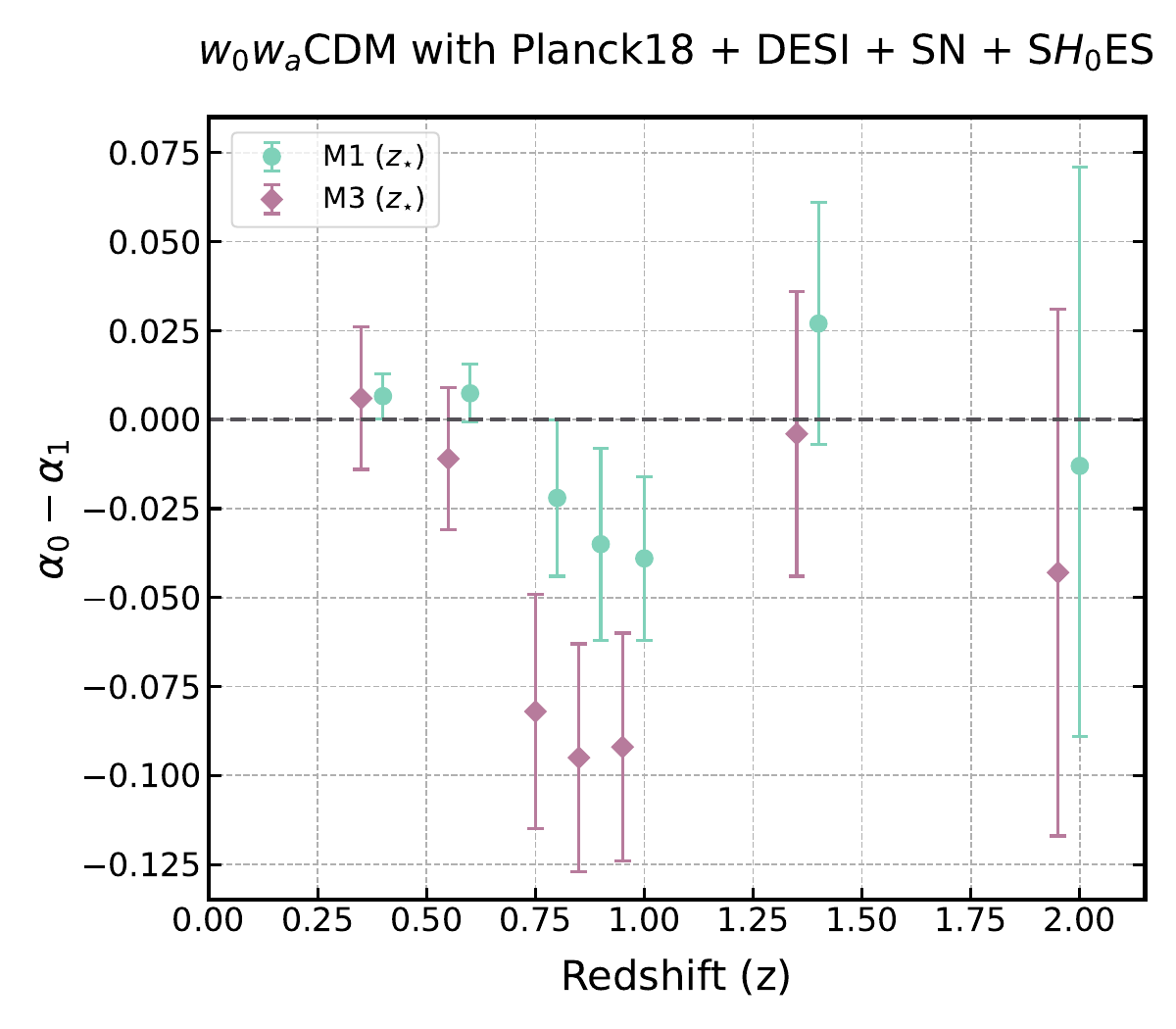}
  \caption{\label{fig:delta_alpha} $\Delta \alpha = \alpha_0 - \alpha_1$ in the M1$(z_*)$ model (round markers) and M3$(z_*)$ model (diamond markers) as a function of $z_*$, assuming a flat $\Lambda$CDM background (left panel) and a $w_0w_a$CDM cosmology (right panel). $\Delta \alpha\neq 0$ suggests redshift evolution is favoured. Note that for visualisation purposes we employ an offset of 0.05 in the redshift of the M3$(z_*)$ data points. The error bars represent the 68\% confidence interval.}
\end{figure*}

To illustrate how changing the transition redshift $z_*$ affects the measured DDR violation, in \cref{fig:delta_alpha} we show $\Delta \alpha = \alpha_0 - \alpha_1$ for M1($z_*$) (sea green) and  M3($z_*$) (magenta), for $z_{*} = \{0.4,0.6,0.8,0.9,1.0,1.4, 2\}$, assuming a flat $\Lambda$CDM (left panel) and $w_0w_a$ background (right panel). $\Delta \alpha\neq 0$ indicates that an evolution of the DDR violation is detected. From the figure, we can see that this occurs in both models when the transition redshift is $0.6 \lesssim z_* \lesssim 1.1$, hinting at evidence for redshift-dependant effects to fit different subsets of the data. This agrees with the results presented in Ref.~\cite{Perivolaropoulos:2023iqj}, as when translated into a magnitude shift according to \cref{eq:Mb_rescale}, we find consistency between both approaches under $\Lambda$CDM at $\sim1\sigma$.

\newpage
\bibliographystyle{apsrev4-1}
\bibliography{bib}

\end{document}